\documentclass{optica-article}

\journal{opticajournal} 

\articletype{Research Article}

\usepackage{graphicx} 
\usepackage{float} 
\usepackage{wrapfig} 
\usepackage[labelfont=bf]{caption} 
\usepackage{caption, subcaption} 

\usepackage{tikz,tikz-3dplot}
\usetikzlibrary{arrows.meta,bending,matrix,positioning,patterns,fit}

\usepackage{bm} 
\usepackage{hyperref} 
\usepackage{lineno} 

\makeatletter
\def\munderbar#1{\underline{\sbox\tw@{$#1$}\dp\tw@\z@\box\tw@}}
\makeatother
\def \opDist {d_{\rm opt}}

 \newcommand{\rmB}{\mathrm{B}}
\newcommand{\rmc}{\mathrm{c}} 
\newcommand{\rmd}{\mathrm{d}} \newcommand{\rmD}{\mathrm{D}}
\newcommand{\rme}{\mathrm{e}} 
 \newcommand{\rmF}{\mathrm{F}}

 \newcommand{\rmI}{\mathrm{I}}

 \newcommand{\rmN}{\mathrm{N}}
 \newcommand{\rmO}{\mathrm{O}}
\newcommand{\rmP}{\mathrm{P}}
 \newcommand{\rmQ}{\mathrm{Q}}
 
\newcommand{\rms}{\mathrm{s}} \newcommand{\rmS}{\mathrm{S}}
\newcommand{\rmt}{\mathrm{t}} \newcommand{\rmT}{\mathrm{T}}

\newcommand{\bsa}{\boldsymbol{a}} \newcommand{\bsA}{\boldsymbol{A}}
\newcommand{\bsb}{\boldsymbol{b}} 
 \newcommand{\bsC}{\boldsymbol{C}}

\newcommand{\bse}{\boldsymbol{\munderbar{e}}}

 \newcommand{\bsI}{\boldsymbol{I}}

\newcommand{\bsm}{\boldsymbol{m}} 
\newcommand{\bsn}{\boldsymbol{n}} 
\newcommand{\bso}{\boldsymbol{\munderbar{o}}} \newcommand{\bsO}{\boldsymbol{O}}
\newcommand{\bsp}{\boldsymbol{p}} \newcommand{\bsP}{\boldsymbol{P}}
\newcommand{\bsq}{\boldsymbol{q}} 
\newcommand{\bsr}{\boldsymbol{\munderbar{r}}} 
\newcommand{\bss}{\boldsymbol{\munderbar{s}}} 
\newcommand{\bst}{\boldsymbol{\munderbar{t}}} 
 
\newcommand{\bsv}{\boldsymbol{\munderbar{v}}} 
 
\newcommand{\bsx}{\boldsymbol{x}} 
\newcommand{\bsy}{\boldsymbol{y}} 
\newcommand{\bsz}{\boldsymbol{z}}

 \newcommand{\bfP}{\mathbf{P}}

 \newcommand{\bfS}{\mathbf{S}}

\newcommand{\bseta}       {\boldsymbol{\eta}}

\newcommand{\bzero}       {\boldsymbol{0}}

 \newcommand{\mcA}{\mathcal{A}}
 \newcommand{\mcB}{\mathcal{B}}
 \newcommand{\mcC}{\mathcal{C}}

 \newcommand{\mcM}{\mathcal{M}}

 \newcommand{\mcP}{\mathcal{P}}
 
 \newcommand{\mcR}{\mathcal{R}}
 \newcommand{\mcS}{\mathcal{S}}
 \newcommand{\mcT}{\mathcal{T}}
 \newcommand{\mcU}{\mathcal{U}}

 \newcommand{\mcX}{\mathcal{X}}
 \newcommand{\mcY}{\mathcal{Y}}

\newcommand{\dfds}[2]{ \frac{\rmd {#1}}{\rmd {#2}} }
\newcommand{\tdfds}[2]{ \tfrac{\rmd {#1}}{\rmd {#2}} }
\newcommand{\pdfds}[2]{ \frac{\partial {#1}}{\partial {#2}} }
\newcommand{\tpdfds}[2]{ \tfrac{\partial {#1}}{\partial {#2}} }

\newcommand{ \dotproduct }   { \boldsymbol{\cdot} }
\newcommand{ \crossproduct } { \boldsymbol{\times} }

\newcommand{\dR}{ \mathbb{R} }

\def \aa  {\mathfrak{a}}

\usepackage{graphicx} 
\usepackage{cleveref}
\usepackage{ulem}
\usepackage{pgfplots}
\usepackage{pgfmath}
\usetikzlibrary{arrows.meta, 3d, calc, decorations.pathreplacing}
\usepackage{caption}
\usepackage{tikz,tikz-3dplot,bm}
\usetikzlibrary{arrows.meta,bending,matrix,positioning,patterns,fit}

\pgfplotsset{compat=1.18}
\begin{document}
\title{Inverse methods for freeform optical design}
\author{J. H. M. ten Thije Boonkkamp,\authormark{1} K. Mitra,\authormark{1} M. J. H. Anthonissen,\authormark{1} L. Kusch,\authormark{1} P. A. Braam,\authormark{1} and W. L. IJzerman\authormark{1,2}}
\address{\authormark{1}CASA, Department of Mathematics and Computer Science, Eindhoven University of Technology, P.O. Box 513, 5600 MB Eindhoven, The Netherlands\\
\authormark{2}Signify Research, High Tech Campus 7, 5656 AE Eindhoven, The Netherlands}
\email{\authormark{*}j.h.m.tenthijeboonkkamp@tue.nl}

\begin{abstract*}
We present a systematic derivation of three mathematical models of increasing complexity for optical design, based on Hamilton's characteristic functions and conservation of luminous flux, and briefly explain the connection with the mathematical theory of optimal transport. We outline several iterative least-squares solvers for our models and demonstrate their performance for a few challenging problems.
\end{abstract*}

\section{Introduction}
Nowadays, traditional incandescent light bulbs have largely been replaced by LED lamps, mainly because LEDs are very energy efficient and have long lifetimes. However, to create a desired light output with LEDs, optical systems are required. Therefore, in the illumination optics industry a lot of research is carried out on the design of optical systems for a myriad of applications.

The industry standard in optical design is (quasi-)Monte Carlo ray tracing. These are forward methods, which compute the target light distribution for a given source, typically LED, and a \textit{given} optical system, tracing a large collection of rays from source to target. Convergence of ray tracing is slow, and requires billions  of rays to accurately compute the target distribution; see \cite[Chapter 3]{Filosa2018}. Moreover, to realize the \textit{desired} target distribution, these methods have to be embedded in an iterative procedure, adjusting system parameters, which makes them inefficient.

A promising alternative are inverse optical design methods.  These methods compute the shape/location of the optical surface(s) in one go, given a source distribution and a \textit{desired} target distribution. The surfaces are referred to as freeform since they have no imposed symmetries. Mathematical models for inverse methods are based on the principles of geometrical optics and conservation of luminous flux. These models contain the following: a geometrical equation describing the shape/location of the optical surface(s), equations for the optical map, which expresses target coordinates as a function of source coordinates, and a conservation law of luminous flux. The geometrical equation has multiple solutions. Combining one of the equations for the optical map with the balance equation for luminous flux, we can derive a Jacobian equation for the optical map. This equation has to be supplemented with the unconventional transport boundary condition.

In mathematics, the abstract theory of optimal transport is the correct framework to formulate optical design problems. The subject of optimal transport theory is to compute a transport plan (mapping) that transforms a given density function into another one, minimizing the transportation cost and subject to a(n) (integral) conservation law; see \cite{Santambrogio2015}. One of the governing equations is precisely the geometrical equation, containing a so-called cost function in the right hand side. The transportation cost is then a cost functional (an integral of the cost function). The analogy with optical design is obvious, the two density functions correspond to the source and target distributions, the transport plan to the optical map, the cost functional can be associated with the optical path length and the conservation law applies to the luminous flux.

Based on the analogy with optimal transport theory, we can distinguish three mathematical models of increasing complexity. First, in the simplest model the geometrical equation contains a quadratic cost function, and the Jacobian equation is the standard Monge-Amp\`{e}re (MA) equation, which is a second-order, nonlinear, elliptic partial differential equation (PDE); \cite{Gutierrez2016}. In the next model, the cost function is no longer quadratic and the Jacobian equation is a generalized Monge-Amp\`{e}re (GMA) equation. Finally, the most complicated model does not allow a description in terms of a cost function anymore, instead a so-called generating function is required. The Jacobian equation is now a generated Jacobian (GJ) equation.

Inverse methods require the numerical solution of a nonlinear elliptic PDE of Monge-Amp\`{e}re type, for which a variety of methods exist. We just mention a few publications and refer to \cite[Chapter 5]{Romijn2021} for a detailed overview. The publication that inspired us to develop our numerical solvers is by \cite{Caboussat2013}. In this paper the authors present a least-squares method to solve the Dirichlet problem for the standard MA equation. Prins adjusted their method to include the transport boundary condition; see \cite{Prins2015}. In a series of papers Yadav and Romijn extended the method for the GMA equation, corresponding to a non-quadratic cost function; see \cite{Yadav2019b}, \cite{Romijn2019}, \cite{Romijn2020}. Romijn further extended our least-squares methods to deal with the GJ equation; see \cite{Romijn2021a} and \cite{Romijn2021b}.
On the other hand, based on the work of \cite{Caffarelli2008}, geometrical approaches of constructing optical surfaces with paraboloids have been investigated by \cite{Kitawaga2019} and \cite{Merigot2021} for the GMA equation, and by \cite{Gallouet2022} for the GJ equation. Finally, adhering to a PDE perspective, numerical algorithms for solving the standard MA equation with transport boundary condition have been investigated by \cite{Benamou2014} using finite differences, \cite{Kawecki2018} using finite elements, and \cite{Froese2012} through an iterative scheme, to name some examples.

For the cost function models we have developed a two-stage algorithm, i.e., we first compute the optical map and subsequently the shape/location of the optical surface(s). For both stages we apply a least-squares method. For the generating function model, the optical map and the optical surface(s) have to be computed
simultaneously, also in a least-squares fashion.

In this paper we present a systematic derivation of all models and outline numerical simulation methods.
The content is then the following. First, in Section \ref{s2:Hamilton} we introduce Hamilton's characteristic functions needed to derive the geometrical equations. In Section \ref{s3:geometry} we give an outline of optimal transport theory, needed to select a uniquely defined solution of the geometrical equation. A generic conservation law for luminous flux is presented in Section \ref{s4:EC}. Next, in Section \ref{s5:hierarchy} we apply the results of the previous sections to present a hierarchy of models based on three example systems. A concise description of our numerical methods is presented in Section \ref{s6:LS}. To demonstrate the performance of our methods, we present a few challenging examples in Section \ref{s7:Examples}. Concluding remarks are given in Section \ref{s8:Conclusions} and the paper is concluded with a nomenclature list.

\section{Hamilton's characteristic functions }
\label{s2:Hamilton}

In this section we introduce Hamilton's characteristic functions which are needed for the derivation of the geometrical equations in \Cref{s5:hierarchy}; see, e.g., \cite[pp. 94 - 107]{Luneburg1966}, \cite[Section 4.1]{Born1975} for a more detailed account. Our starting point is the eikonal equation given by
\begin{equation} \label{s2:eikonal}
  |\nabla \varphi| = n,
\end{equation}
where $\varphi = \varphi(\bsr)$ is the phase of an electromagnetic wave and $n = n(\bsr)$ is the refractive index field, as functions of the (three-dimensional) position vector (or coordinates) $\bsr \in \dR^{3}$ of a typical point. A surface $\varphi(\bsr) = \mathrm{Const}$ is a wavefront. We denote three-dimensional vectors in bold font with an underbar (\,$\underbar{\;\;}$\,), to distinguish them from two-dimensional vectors, which are only written in bold font. Equation (\ref{s2:eikonal}) describes free propagation of light waves and no longer holds when a light wave is reflected or refracted at an optical surface. The eikonal equation (\ref{s2:eikonal}) is a first-order nonlinear PDE, which has the characteristic strip equations
\begin{equation} \label{s2:char_eq}
  \dfds{\bsr}{s} = \frac{1}{n} \nabla \varphi, \quad \dfds{\varphi}{s} = n, \quad \dfds{ }{s}(\nabla \varphi) = \nabla n,
\end{equation}
see, e.g., \cite{Kevorkian1990}. The curves $\mcC:\bsr = \bsr(s)$, parameterized by the arc length $s$, are the characteristics of (\ref{s2:eikonal}). The system of ordinary differential equations in (\ref{s2:char_eq}) determines the location of the characteristics and the solution along the characteristics, from which the solution of (\ref{s2:eikonal}) can be reconstructed. From the first equation in (\ref{s2:char_eq}) we conclude that the characteristics are orthogonal trajectories to the wavefronts, i.e., characteristics are light rays; see, e.g., \cite[p. 114]{Born1975}.

Hamilton's characteristic functions are related to the optical path length of a ray, which we therefore introduce next. We assume that $n(\bsr)$ is piecewise continuous, with jumps at lens surfaces, consequently, rays are continuous and consist of piecewise differentiable curve-segments; cf. (\ref{s2:char_eq}).
In this setting, we define the optical path length (OPL) between the points $\rmP_{1}$ (position vector $\bsr = \bsr_{1}$) and $\rmP_{2}$ (position vector $\bsr = \bsr_{2})$, connected by a light ray along the curve $\mcC$, as
\begin{align}\label{s2:opDist}
  [\rmP_{1},\rmP_{2}] = \opDist( \bsr_{1},\bsr_{2} ):= \int_{\mcC} n(\bsr(s))\, \rmd s.
\end{align}
A consequence of Fermat's principle is that the optical path length $[\rmP_{1},\rmP_{2}]$ is stationary with respect to variations of the curve $\mcC$; see, e.g., \cite[p. 740]{Born1975}.

Consider an optical system consisting of a source, located in a plane $z = z_{\rms}$ (source plane), emitting light which via a sequence of reflectors and/or lenses arrives at a plane $z = z_{\rmt}$ (target plane).
The $z$-axis is referred to as the optical axis. In this paper, we use the subscripts $\rms$ and $\rmt$ for variables/vectors to denote their value at source and target, respectively. Let $\hat{\bsv} = ( v_{1}\, v_{2}\, v_{3} )^{\rmT} = \rmd \bsr / \rmd s$ denote the unit tangent vector of a ray, thus $\nabla \varphi = n \hat{\bsv}$; cf. Eq. (\ref{s2:char_eq}). Unit vectors are denoted with a hat ($\,\hat{ }\,$).
Let $\rmQ_{\rms}$ (position vector $\bsr = \bsr_{\rms}$) and $\rmQ_{\rmt}$ (position vector $\bsr = \bsr_{\rmt}$) be two points on the source and target plane, respectively, then from (\ref{s2:opDist}) we obtain
\begin{align}\label{s2:distGrad}
  [ \rmQ_{\rms},\rmQ_{\rmt} ] = \opDist( \bsr_{\rms},\bsr_{\rmt} ) = \int_{\mcC} n \hat{\bsv} \dotproduct \hat{\bsv}\, \rmd s =
  \int_{\mcC} \nabla \varphi \dotproduct \rmd \bsr = \varphi( \bsr_{\rmt} ) - \varphi( \bsr_{\rms} ).
\end{align}
Note that reversing the direction of integration yields the symmetry property of $\opDist$, i.e.,
\begin{align}\label{s2:distSym}
  \opDist( \bsr_{\rmt},\bsr_{\rms} ) = \int_{-\mcC} (-n \hat{\bsv} ) \dotproduct (-\hat{\bsv})\, \rmd s =
  \int_{-\mcC} -\nabla \varphi \dotproduct \rmd \bsr = -\varphi( \bsr_{\rms} ) + \varphi( \bsr_{\rmt} )
  = \opDist( \bsr_{\rms},\bsr_{\rmt} ),
\end{align}
as one would expect from the definition in (\ref{s2:opDist}).
Observe that $\opDist$ is a metric since $\opDist( \bsr_{\rms},\bsr_{\rmt} ) \geq 0$, with equality only if $\bsr_{\rms} = \bsr_{\rmt}$, and it satisfies the triangle inequality due to Fermat's principle. The symmetry of $\opDist$ is stated in \eqref{s2:distSym}.

For ease of presentation, we use the shorthand notation $\hat{\bsv}_{\rms} = \hat{\bss}$ and $\hat{\bsv}_{\rmt} = \hat{\bst}$.
Straightforward differentiation of the expression for $\opDist( \bsr_{\rms},\bsr_{\rmt} )$ in (\ref{s2:distGrad}) gives
\begin{equation} \label{s2:gradients}
  \nabla_{\rms} \opDist( \bsr_{\rms},\bsr_{\rmt} ) = -\nabla \varphi( \bsr_{\rms} ) = -n_{\rms} \hat{\bss}, \quad
  \nabla_{\rmt} \opDist( \bsr_{\rms},\bsr_{\rmt} ) = \nabla \varphi( \bsr_{\rmt} ) = n_{\rmt} \hat{\bst},
\end{equation}
where $\nabla_{\rms}$ denotes the gradient with respect to the source coordinates $\bsr_{\rms}$ and likewise for $\nabla_{\rmt}$. From these relations we readily conclude that
\[
  |\nabla_{\rms} \opDist| = n_{\rms}, \quad |\nabla_{\rmt} \opDist| = n_{\rmt},
\]
i.e., $\opDist$ satisfies the eikonal equation in both source and target coordinates.

A ray is completely determined by its position and direction coordinates.
Position coordinates $\bsq \in \dR^{2}$ are defined as the projection on a plane $z = \mathrm{Const}$ of the position vector $\bsr$ of a typical point on the ray. Position coordinates in the source and target planes are denoted by $\bsq_{\rms}$ and $\bsq_{\rmt}$, respectively. Thus $\rmQ_{\rms}$, with position vector $\bsr_{\rms} = ( \bsq_{\rms},z_{\rms} )$, will denote the point at which a light ray is emitted from the source plane, and $\rmQ_{\rmt}$, with position vector $\bsr_{\rmt} = ( \bsq_{\rmt},z_{\rmt})$, as the point at which it arrives at the target plane. Next, selecting the first two components from the equation $n \tdfds{\bsr}{s} = n \hat{\bsv} = \nabla \varphi$, we introduce the momentum vector $\bsp \in \mathbb{R}^{2}$ as
\begin{equation} \label{s2:momentum}
  \bsp := n \dfds{\bsq}{s} = n \begin{pmatrix} v_{1} \\ v_{2} \end{pmatrix} = \begin{pmatrix} \tpdfds{\varphi}{q_{1}} \\[0.2cm] \tpdfds{\varphi}{q_{2}} \end{pmatrix} =: \pdfds{\varphi}{\bsq}.
\end{equation}
Thus, the momentum vector is the projection of  $\nabla \varphi(\bsx)$ on a plane $z = \mathrm{Const}$ and determines the direction of a light ray. Momentum vectors at source and target are denoted with $\bsp_{\rms}$ and $\bsp_{\rmt}$, respectively. The variables $\bsq = \bsq(z)$ and $\bsp = \bsp(z)$ are referred to as phase-space coordinates and together constitute phase space.

In the remainder of this paper, we assume that $n(\bsr)$ is piecewise constant, with possibly discontinuities at optical interfaces, and consequently a light ray consists of piecewise straight line segments. Below, we define four characteristic functions which relate the optical path length to phase-space coordinates at source and target.
\vspace{0.1cm}

\noindent
\textbf{Point characteristic.} As a function of the position coordinates $\bsq_{\rms}$ and $\bsq_{\rmt}$, the point characteristic $V$ is the optical path length of the ray connecting the points $\rmQ_{\rms}$ and $\rmQ_{\rmt}$, i.e.
\begin{equation} \label{s2:V_char}
  V( \bsq_{\rms},\bsq_{\rmt} ) := \opDist( (\bsq_{\rms},z_{\rms}),(\bsq_{\rmt},z_{\rmt}) ) =
  [ \rmQ_{\rms},\rmQ_{\rmt} ].
\end{equation}
From (\ref{s2:distGrad}) we conclude that $V( \bsq_{\rms},\bsq_{\rmt} ) = \varphi( \bsq_{\rmt},z_{\rmt} ) - \varphi( \bsq_{\rms},z_{\rms} )$. Combining this relation with the equations in (\ref{s2:gradients}), selecting the first two components in both equations, we find for the directional derivatives $\tpdfds{V}{\bsq_{\rms}}$ and $\tpdfds{V}{\bsq_{\rmt}}$ the following expressions
\begin{equation} \label{s2:directional}
  \pdfds{V}{\bsq_{\rms}} = -n_{\rms} \begin{pmatrix} s_{1} \\ s_{2} \end{pmatrix} = -\bsp_{\rms}, \quad \pdfds{V}{\bsq_{\rmt}} = n_{\rmt} \begin{pmatrix} t_{1} \\ t_{2} \end{pmatrix} = \bsp_{\rmt};
\end{equation}
cf. Eq. \eqref{s2:momentum}. Typically, Eqs. \eqref{s2:V_char} and \eqref{s2:directional} are used to derive a geometrical equation for optical systems described in terms of the position coordinates $\bsq_{\rms}$ and $\bsq_{\rmt}$. An example is the parallel-to-near-field reflector which we discuss in Section \ref{s5.3:Pa2NFR}.
\vspace{0.1cm}

\begin{figure}[!ht]
  \centering
\begin{subfigure}[b]{0.48\textwidth}
    \includegraphics[width = 1.0\textwidth,trim={0cm 0cm 0cm 0cm},clip ]{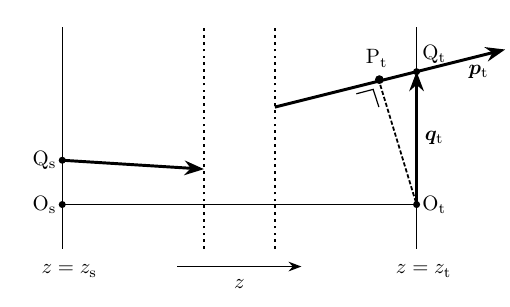}
    \caption{ Mixed characteristic $W(\bsq_{\rms},\bsp_{\rmt})$ of the first kind;\\ $\sigma_{\rmt} > 0$. }
    \label{Fig:interpretationW1}
\end{subfigure}
\hfill
\begin{subfigure}[b]{0.48\textwidth}
    \includegraphics[width = 1.0\textwidth,trim={0cm 0cm 0cm 0cm},clip ]{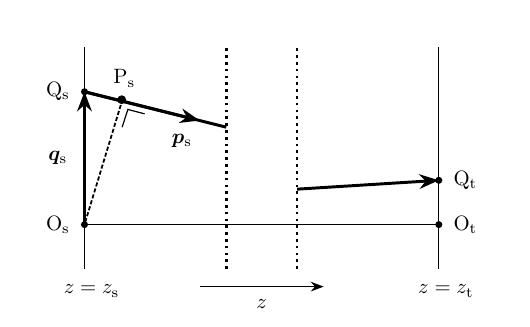}
    \caption{ Mixed characteristic $W^{\ast}(\bsp_{\rms},\bsq_{\rmt})$ of the second kind;\\ $\sigma_{\rms} < 0$. }
    \label{Fig:interpretationW2}
\end{subfigure}
\caption{ Interpretation of the mixed characteristics of the first and second kind. }
\label{Fig:interpretationW}
\end{figure}

\noindent
\textbf{Mixed characteristic of the first kind.} Let a light ray connect the points
$\rmQ_{\rms}( \bsq_{\rms}, z_{\rms} )$ and $\rmQ_\rmt( \bsq_{\rmt},z_{\rmt} )$ with momentum $\bsp_{\rms}$ and $\bsp_{\rmt}$, respectively, then the mixed characteristic of the first kind $W$ is defined by
\begin{equation} \label{s2:W_char}
  W( \bsq_{\rms},\bsp_{\rmt} ) := V( \bsq_{\rms},\bsq_{\rmt} ) - \bsq_{\rmt} \dotproduct \bsp_{\rmt}.
\end{equation}
Straightforward differentiation gives $\partial W / \partial \bsp_{\rms} = \bzero$ and $\partial W / \partial \bsq_{\rmt} = \bzero$, using Eq. \eqref{s2:directional}, consequently $W = W( \bsq_{\rms},\bsp_{\rmt} )$ indeed. Moreover, we readily verify that
\begin{align}\label{s2:directional1}
\frac{\partial W}{\partial \bsq_{\rms}} = - \bsp_{\rms}, \quad \frac{\partial W}{\partial \bsp_{\rmt}} = - \bsq_{\rmt}.
\end{align}
$W$ can be interpreted as a modification of $[ \rmQ_{\rms},\rmQ_{\rmt} ]$, which we show as follows.
Assume that $z_{\rmt} > z_{\rms}$ and define the point $\rmP_{\rmt}$ as the intersection of the ray segment arriving at $\rmQ_{\rmt}$, parameterized by $\bsr(\lambda) = \bsr_{\rmt} + \lambda \hat{\bst}$, and the plane passing through the origin $\rmO_{\rmt}(\bzero,z_{\rmt})$ of the target plane (position vector $\bso_{\rmt}$) perpendicular to this ray, given by the normal equation $(\bsr - \bso_{\rmt}) \dotproduct \hat{\bst} = 0$; see Figure \ref{Fig:interpretationW}.
The intersection point is given by $\lambda = \lambda(\rmP_{\rmt}) = -(\bsr_{\rmt}-\bso_{\rmt}) \dotproduct \hat{\bst}$. Since the third component of $\bsr_{\rmt} - \bso_{\rmt}$ vanishes, this can be rewritten as $\lambda(\rmP_{\rmt}) = -\bsq_{\rmt} \dotproduct \begin{pmatrix} t_{1} \\ t_{2} \end{pmatrix} = - \bsq_{\rmt} \dotproduct \bsp_{\rmt} / n_{\rmt}$; cf. Eq. (\ref{s2:momentum}) or Eq. (\ref{s2:directional}). For the $z$-component of $\rmP_{\rmt}$ we have $\delta z_{\rmt} := z(\rmP_{\rmt}) - z_{\rmt} = \lambda(\rmP_{\rmt}) t_{3}$ with $t_{3} > 0$ since the ray is directed towards the target plane ($z_{\rmt} > z_{\rms}$). From this we conclude that $\lambda(\rmP_{\rmt}) > 0$ if $\delta z_{\rmt} > 0$, i.e., $\rmP_{\rmt}$ is located behind the target plane as seen from the source, and $\lambda(\rmP_{\rmt}) < 0$ otherwise. Obviously, the distance $d(\rmQ_{\rmt},\rmP_{\rmt}) = \pm \lambda(\rmP_{\rmt})$ since $\hat{\bst}$ is a unit vector. If $\delta z_{\rmt} > 0$, then $\lambda(\rmP_{\rmt}) > 0$, implying that $d( \rmQ_{\rmt},\rmP_{\rmt} ) = \lambda(\rmP_{\rmt})$ and $\bsq_{\rmt} \dotproduct \bsp_{\rmt} = - n_{\rmt} d( \rmQ_{\rmt},\rmP_{\rmt} ) = -[\rmQ_{\rmt},\rmP_{\rmt} ]$; cf. Eq. (\ref{s2:opDist}). Analogously, if $\delta z_{\rmt} < 0$ then $\bsq_{\rmt} \dotproduct \bsp_{\rmt} = [ \rmQ_{\rmt},\rmP_{\rmt} ]$. Finally, we can easily verify that $\sigma_{\rmt} := \mathrm{sgn}( \bsq_{\rmt} \dotproduct \bsp_{\rmt} ) = - \mathrm{sgn}( \delta z_{\rmt} )$, thus $W$ can be interpreted as
\[
  W( \bsq_{\rms},\bsp_{\rmt} ) = [ \rmQ_{\rms},\rmQ_{\rmt} ] - \sigma_{\rmt} [ \rmQ_{\rmt},\rmP_{\rmt} ].
\]
The characteristic function $W$ is convenient to describe optical systems in terms of $\bsq_{\rms}$ and $\bsp_{\rmt}$. An example is the parallel-to-far-field reflector as described in \Cref{s5.1:Pa2FFR}.

\vspace{0.1cm}

\noindent
\textbf{Mixed characteristic of the second kind.} Under the same setting, the mixed characteristic of the second kind $W^{\ast}$ is defined as
\begin{equation} \label{s2:Wast_char}
  W^{\ast}( \bsp_{\rms},\bsq_{\rmt} ) := V( \bsq_{\rms},\bsq_{\rmt} ) + \bsq_{\rms} \dotproduct \bsp_{\rms}.
\end{equation}
In this case $\partial W^{\ast} / \partial \bsq_{\rms} = \bzero$, using Eq. \eqref{s2:directional}, and $\partial W^{\ast} / \partial \bsp_{\rmt} = \bzero$, so $W^{\ast}$ is a function of $\bsp_{\rms}$ and $\bsq_{\rmt}$ indeed. Furthermore, straightforward differentiation gives
\begin{align*}
\frac{\partial W^{\ast}}{\partial \bsp_{\rms}} = \bsq_{\rms}, \quad \frac{\partial W^{\ast}}{ \partial \bsq_{\rmt}} = \bsp_{\rmt}.
\end{align*}
Also $W^{\ast}$ can be interpreted as a modified OPL. We assume again that $z_{\rmt} > z_{\rms}$ and
introduce the point $\rmP_{\rms}$ as the intersection of the ray segment emitted from $\rmQ_{\rms}$, parameterized by $\bsr(\lambda) = \bsr_{\rms} + \lambda \hat{\bss}$, and the plane passing through the origin $\rmO_{\rms}( \bzero,z_{\rms} )$ of the source plane (position vector $\bso_{\rms}$) perpendicular to this ray, given by $(\bsr - \bso_{\rms}) \dotproduct \hat{\bss} = 0$; see Figure \ref{Fig:interpretationW}. Analogous to the previous derivation we obtain $\lambda = \lambda(\rmP_{\rms}) = -( \bsr_{\rms} - \bso_{\rms} ) \dotproduct \hat{\bss} = - \bsq_{\rms} \dotproduct \begin{pmatrix} s_{1} \\ s_{2} \end{pmatrix} = - \bsq_{\rms} \dotproduct \bsp_{\rms} / n_{\rms}$. The $z$-component of $\rmP_{\rms}$ is given by $z(\rmP_{\rms}) = z_{\rms} + \lambda(\rmP_{\rms}) s_{3}$ with $s_{3} > 0$. Thus, if $\delta z_{\rms} := z(P_{\rms}) - z_{\rms} > 0$ then $\lambda(\rmP_{\rms}) > 0$, otherwise, $\lambda(\rmP_{\rms}) < 0$. Following the same line of reasoning as before, we find that $\bsq_{\rms} \dotproduct \bsp_{\rms} = \sigma_{\rms} [\rmQ_{\rms},\rmP_{\rms} ]$ with $\sigma_{\rms} = \mathrm{sgn}( \bsq_{\rms} \dotproduct \bsp_{\rms} )$. Therefore, the second mixed characteristic can be interpreted as
\[
  W^{\ast}( \bsp_{\rms},\bsq_{\rmt} ) = [ \rmQ_{\rms},\rmQ_{\rmt} ] + \sigma_{\rms} [ \rmQ_{\rms},\rmP_{\rms} ].
\]
Note that $W^{\ast}( -\bsp_{\rmt},\bsq_{\rms} ) = W (\bsq_{\rms},\bsp_{\rmt} )$, i.e., the mixed characteristic of the second kind of the ray in reversed direction equals the mixed characteristic of the first kind of the original ray.
The function $W^{\ast}$ can be used to characterize optical systems in terms of $\bsp_{\rms}$ and $\bsq_{\rmt}$.
An example is the point-to-near-field reflector.
\vspace{0.1cm}

\noindent
\textbf{Angular characteristic.} The angular characteristic $T$ is another modification of $[\rmQ_{\rms},\rmQ_{\rmt}]$ and is a function of the momenta $\bsp_{\rms}$ and $\bsp_{\rmt}$. Its definition and interpretation (for $z_{\rmt} > z_{\rms}$) are given by
\begin{equation} \label{s2:T_char}
\begin{split}
   T( \bsp_{\rms},\bsp_{\rmt} ) &:= V( \bsq_{\rms},\bsq_{\rmt} ) + \bsq_{\rms} \dotproduct \bsp_{\rms} - \bsq_{\rmt} \dotproduct \bsp_{\rmt}\\
  &= [ \rmQ_{\rms},\rmQ_{\rmt} ] + \sigma_{\rms} [ \rmQ_{\rms},\rmP_{\rms} ]
   - \sigma_{\rmt} [ \rmQ_{\rmt},\rmP_{\rmt} ].
\end{split}
\end{equation}
From Eq. (\ref{s2:directional}) and Eq. (\ref{s2:T_char}) we readily verify that $\partial T / \partial \bsq_{\rms} = \bzero$, $\partial T / \bsq_{\rmt} = \bzero$, implying that $T = T( \bsp_{\rms},\bsp_{\rmt} )$ indeed. Moreover, it is evident that
\[
  \frac{\partial T}{\partial \bsp_{\rms}} = \bsq_{\rms}, \quad \frac{\partial T}{ \partial \bsp_{\rmt} }= - \bsq_{\rmt}.
\]
Note that $T( -\bsp_{\rmt},-\bsp_{\rms} ) = T( \bsp_{\rms},\bsp_{\rmt} )$, i.e., the angular characteristic is invariant if the direction of the ray is reversed. This function is used to describe optical systems in terms of $\bsp_{\rms}$ and $\bsp_{\rmt}$, such as the point-to-far-field lens; see Section \ref{s5.2:Po2FFL}.


\section{Geometrical description of optical systems: an optimal transport point of view}
\label{s3:geometry}

To derive a geometrical description of an optical system, i.e., an algebraic relation defining the location and shape of the optical surface(s), we have to evaluate one of the four Hamilton's characteristic functions. However, this geometrical equation does not have a unique solution. In this section we discuss how to select a uniquely defined solution from this equation. Subsequently, we derive an equation for the so-called optical map, which relates source and target coordinates of a ray.

First we introduce some notation. We denote with $\mcS$ and $\mcT$ the (physical) source and target domain, respectively. Source and target domain are parameterized by $\bsx \in \mcX \subset \dR^{2}$ and $\bsy \in \mcY \subset \dR^{2}$, respectively, i.e., $\mcX$ and $\mcY$ are the parameter domains for source and target. We assume that $\mcX$ and $\mcY$ are compact, i.e., bounded and closed, and consequently any continuous function defined on one of these domains assumes a minimum and a maximum. Key in the following discussion is the optical map $\bsm: \mcX \to \mcY$, defining how a point on the source domain (specified by $\bsx$) is connected via a light ray with a point on the target domain (specified by $\bsy = \bsm(\bsx)$). Given source and target domain, the optical map has to satisfy the condition $\mcY = \bsm(\mcX)$, i.e., $\mcY$ is the image of $\mcX$ under the mapping $\bsm$.

We consider optical systems having one or two freeform optical surfaces, i.e., surfaces without any symmetry.
Consider an arbitrary ray connecting source and target, for which $\bsy = \bsm(\bsx)$.
Evaluation of the appropriate Hamilton's characteristic function for such a ray leads to either one of the two algebraic equations
\begin{subequations} \label{s3:geom_eq}
\begin{align}
  u_{1}(\bsx) + u_{2}(\bsy) &= c(\bsx,\bsy), \label{s3:geom_eqa}\\
  u_{2}(\bsy) &= H( \bsx,\bsy,u_{1}(\bsx) ), \label{s3:geom_eqb}
\end{align}
\end{subequations}
which we refer to as geometrical equations,
where $u_{1}(\bsx)$ defines the location of (one of) the optical surface(s) and $u_{2}(\bsy)$ is either an auxiliary variable or defines the location of the other surface. We like to emphasize that in Eqs. (\ref{s3:geom_eq}) source and target coordinates are related via $\bsy = \bsm(\bsx)$.
In Section \ref{s5:hierarchy} we will derive geometrical equations for several optical systems.
Note that in Eq. (\ref{s3:geom_eqa}) source and target coordinates are separated in the left-hand side. This equation occurs in the theory of optimal transport, where $u_{1}(\bsx)$ and $u_{2}(\bsy)$ are referred to as Kantorovich potentials and $c(\bsx,\bsy)$ as the cost function; see, e.g., \cite{Santambrogio2015} for a rigorous mathematical account. In Eq. (\ref{s3:geom_eqb}) source and target coordinates are no longer separated. It will turn out that for fixed $\bsx$ and $\bsy$, the function $H( \bsx,\bsy,\cdot )$ has a unique inverse $H^{-1}( \bsx,\bsy,\cdot ) = G( \bsx,\bsy,\cdot )$, referred to as the generating function; see \cite{Guillen2019}. Thus the following holds
\begin{equation} \label{s3:inverse}
  \forall \bsx \in \mcX\, \forall \bsy \in \mcY \quad \big( u_{2}(\bsy) = H( \bsx,\bsy,u_{1}(\bsx) ) \quad \Longleftrightarrow \quad
    u_{1}(\bsx) = G( \bsx,\bsy,u_{2}(\bsy) ) \big).
\end{equation}
Obviously, Eq. (\ref{s3:geom_eqa}) is a special case of Eq. (\ref{s3:geom_eqb}) if we choose $H( \bsx,\bsy,w ) = c(\bsx,\bsy) - w$. Nevertheless, for the sake of completeness we discuss both cases separately.
\vspace{0.1cm}

\noindent
\textbf{$c$-convex analysis.}
Equation (\ref{s3:geom_eqa}) has multiple solution pairs $( u_{1}(\bsx), u_{2}(\bsy) )$. From the theory of optimal transport we know that a possible solution of Eq. (\ref{s3:geom_eqa}) is given by
\begin{subequations} \label{s3:max_max_sol}
\begin{align}
  \forall \bsx \in \mcX \quad u_{1}(\bsx) &= \max_{\bsy \in \mcY}\big( c(\bsx,\bsy) - u_{2}(\bsy) \big), \label{s3:max_max_sola}\\
  \forall \bsy \in \mcY \quad u_{2}(\bsy) &= \max_{\bsx \in \mcX}\big( c(\bsx,\bsy) - u_{1}(\bsx) \big), \label{s3:max_max_solb}
\end{align}
\end{subequations}
which implicitly defines the optical map as alternative to a geometrical optics derivation. To be more precise,
for any $\bsx \in \mcX$ its image $\bsy^{\ast} = \bsm(\bsx) \in \mcY$ is the maximizer in (\ref{s3:max_max_sola}). Also, for any $\bsy \in \mcY$ we have that $\bsx^{\ast} \in \mcX$ such that $\bsy = \bsm(\bsx^{\ast})$ is the maximizer in (\ref{s3:max_max_solb}), which is possible since $\mcY = \bsm(\mcX)$. Under certain conditions, to be specified shortly, the maxima in (\ref{s3:max_max_sol}) are unique,  which we henceforth assume.
The solution in (\ref{s3:max_max_sol}) is referred to as a $c$-convex pair; see \cite{Yadav2018}. The proof of the equivalence of (\ref{s3:max_max_sola}) and (\ref{s3:max_max_solb}) proceeds as follows.
Suppose, Eq. (\ref{s3:geom_eqa}) and the expression for $u_{1}(\bsx)$ in Eq. (\ref{s3:max_max_sola}) hold, then we have to derive the expression for $u_{2}(\bsy)$ in Eq. (\ref{s3:max_max_solb}). From (\ref{s3:max_max_sola}) we conclude
\[
  \forall \bsx \in \mcX\, \forall \bsy \in \mcY \quad u_{1}(\bsx) \ge c(\bsx,\bsy) - u_{2}(\bsy),
\]
or equivalently, swapping $u_{1}(\bsx)$ and $u_{2}(\bsy)$,
\[
  \forall \bsx \in \mcX\, \forall \bsy \in \mcY \quad u_{2}(\bsy) \ge c(\bsx,\bsy) - u_{1}(\bsx).
\]
Since the latter inequality holds for all $\bsx \in \mcX$, we can take the maximum over all $\bsx \in \mcX$ to obtain
\[
  \forall \bsy \in \mcY \quad u_{2}(\bsy) \ge \max_{\bsx \in \mcX} \big( c(\bsx,\bsy) - u_{1}(\bsx) \big).
\]
Recall that $\mcY = \bsm(\mcX)$. Let $\bsy \in \mcY$, then $\bsy = \bsm(\tilde{\bsx})$ for some $\tilde{\bsx} \in \mcX$. From (\ref{s3:geom_eqa}) we conclude that
\[
  u_{2}(\bsy) = c(\tilde{\bsx},\bsy) - u_{1}(\tilde{\bsx}) \le \max_{\bsx \in \mcX} \big( c(\bsx,\bsy) - u_{1}(\bsx) \big).
\]
Combining both inequalities we obtain the expression for $u_{2}(\bsy)$ in Eq. (\ref{s3:max_max_solb}) and conclude  that $\tilde{\bsx} = \bsx^{\ast}$ is the (unique) maximizer of $c(\bsx,\bsy) - u_{1}(\bsx)$. Thus, the expression for $u_{1}(\bsx)$ in Eq. (\ref{s3:max_max_sola}) combined with Eq. (\ref{s3:geom_eqa}) implies the expression for $u_{2}(\bsy)$ in Eq. (\ref{s3:max_max_solb}). Conversely, given the expression for $u_{2}(\bsy)$ in Eq. ~(\ref{s3:max_max_solb}) and Eq. (\ref{s3:geom_eqa}) we can derive in a similar manner the expression for $u_{1}(\bsx)$ defined in Eq. (\ref{s3:max_max_sola}).

Alternatively, Eq. (\ref{s3:geom_eqa}) has a $c$-concave solution pair, for which the maxima in Eq. (\ref{s3:max_max_sol}) are replaced by minima, defining another mapping $\bsm$. In either case, a necessary condition is that $(\bsx,\bsm(\bsx))$ is a stationary point of $c(\bsx,\bsy) - u_{1}(\bsx)$ for all $\bsx \in \mcX$, i.e.,
\begin{equation} \label{s3:stationary_point}
  \nabla_{\bsx} c(\bsx,\bsm(\bsx)) - \nabla u_{1}(\bsx) = \bzero,
\end{equation}
where $\nabla_{\bsx} c$ is the gradient of $c$ with respect to the source coordinates $\bsx$. A sufficient condition for the $c$-convex pair (max/max solution) in (\ref{s3:max_max_sol}) is that $\rmD_{\bsx\bsx} c(\bsx,\bsm(\bsx)) - \rmD^{2} u_{1}(\bsx)$ be symmetric negative definite (SND) for all $\bsx \in \mcX$, where $\rmD_{\bsx\bsx} c$ and $\rmD^{2} u_{1}$ are the Hessian matrices of $c$ (with respect to $\bsx$) and $u_{1}$, respectively. The function $c(\cdot,\bsy) - u_{1}$ is then concave, guaranteeing a unique maximum in Eq. (\ref{s3:max_max_solb}). Alternatively, for a $c$-concave pair (min/min solution) the sufficient condition is that $\rmD_{\bsx\bsx} c (\bsx,\bsm(\bsx)) - \rmD^{2} u_{1}(\bsx)$ be symmetric positive definite (SPD) for all $\bsx \in \mcX$. Then $c(\cdot,\bsy) - u_{1}$ is convex and has a unique minimum.

According to the implicit function theorem, the optical map $\bsy = \bsm(\bsx)$ is well defined by Eq. (\ref{s3:stationary_point}), provided the mixed derivative matrix $\rmD_{\bsx\bsy}c = \big( \tfrac{ \partial^{2} c }{ \partial x_{i} \partial y_{j} } \big)$ is regular in $(\bsx,\bsm(\bsx))$ for all $\bsx \in \mcX$, the so-called twist condition.
However, for many optical systems the actual computation of $\bsy = \bsm(\bsx)$ from (\ref{s3:stationary_point}) is quite complicated, if not impossible. Therefore, to derive an equation for the optical map, we differentiate the zero-gradient condition in Eq. (\ref{s3:stationary_point}) to obtain the equation
\begin{equation} \label{s3:matrix_eq}
  \bsC(\bsx,\bsm(\bsx)) \rmD \bsm = \bsP(\bsx,\bsm(\bsx)), \quad \bsC(\bsx,\bsy) = \rmD_{\bsx \bsy} c(\bsx,\bsy),
  \quad \bsP(\bsx,\bsy) = \rmD^{2} u_{1}(\bsx) - \rmD_{\bsx\bsx} c(\bsx,\bsy),
\end{equation}
where $\rmD \bsm = \big( \tfrac{\partial m_{i}}{\partial x_{j}} \big)$ is the Jacobi matrix of the optical map.
We refer to Eq. (\ref{s3:matrix_eq}) as the matrix-Jacobi equation. Note that the matrix $\bsP$ is either SPD, for a $c$-convex, or SND for a $c$-concave solution pair.

Equation (\ref{s3:matrix_eq}) has to be supplemented with the so-called transport boundary condition
\begin{equation} \label{s3:transport_bc}
  \bsm(\partial \mcX) = \partial \mcY,
\end{equation}
stating that the boundary of the source domain $\partial \mcX$ is mapped to the boundary of the target domain $\partial \mcY$. This condition is a consequence of the edge-ray principle of geometrical optics by \cite{Ries2002}, and guarantees that all light emitted from the source arrives at the target.

For some basic optical systems $c(\bsx,\bsy) = \bsx \dotproduct \bsy$, implying that $\bsC = \bsI$ and $\rmD_{\bsx\bsx} c = \bsO$; see Section \ref{s5.1:Pa2FFR}. The $c$-convex solution pair in (\ref{s3:max_max_sola}) reduces to a conjugate pair of Legendre-Fenchel transforms, see e.g., \cite{Romijn2021}, for which the necessary condition (\ref{s3:stationary_point}) reduces to $\bsm(\bsx) - \nabla u_{1}(\bsx) = \bzero$. The sufficient condition for the $c$-convex solution pair is that $\rmD^{2} u_{1}(\bsx)$ is SPD for all $\bsx \in \mcX$, i.e., $u_{1}(\bsx)$ is a convex function. Likewise, for the $c$-concave solution pair $\rmD^{2} u_{1}(\bsx)$ has to be SND for all $\bsx \in \mcX$, and $u_{1}(\bsx)$ is then concave. In both cases $c(\cdot,\bsy) - u_{1}$ has a unique extremum.

In our numerical algorithm, the matrix-Jacobi equation (\ref{s3:matrix_eq}), subject to the transport boundary condition (\ref{s3:transport_bc}), is employed to compute the optical map, and subsequently $u_{1}(\bsx)$ is reconstructed from Eq. (\ref{s3:stationary_point}). If needed, $u_{2}(\bsy)$ is computed from (\ref{s3:geom_eqa}). From $u_{1}(\bsx)$ and possibly $u_{2}(\bsy)$ the shape of the optical surface(s) is computed. A concise account of the numerical algorithm is given in Section \ref{s6:LS}.
\vspace{0.1cm}

\noindent
\textbf{$H$-convex analysis.}
Next, we consider the geometrical equation (\ref{s3:geom_eqb}), which can always be formulated such that
$H_{w}(\bsx,\bsy,\cdot) > 0$ for all $\bsx \in \mcX$ and $\bsy \in \mcY$, implying that $H(\bsx,\bsy,\cdot)$ has an inverse. Also Eq. (\ref{s3:geom_eqb}) has multiple solution pairs $(u_{1}(\bsx), u_{2}(\bsy))$. In analogy with (\ref{s3:max_max_sol}) a possible solution reads
\begin{subequations} \label{s3:max_min_sol}
\begin{align}
  \forall \bsx \in \mcX \quad u_{1}(\bsx) &= \max_{\bsy \in \mcY} G( \bsx,\bsy,u_{2}(\bsy) ), \label{s3:max_min_sola}\\
  \forall \bsy \in \mcY \quad u_{2}(\bsy) &= \min_{\bsx \in \mcX} H( \bsx,\bsy,u_{1}(\bsx) ), \label{s3:max_min_solb}
\end{align}
\end{subequations}
implicitly defining the optical map as follows: $\bsy^{\ast} = \bsm(\bsx) \in \mcY$ is the maximizer in (\ref{s3:max_min_sola}) and $\bsx^{\ast} \in \mcX$ such that $\bsy = \bsm(\bsx^{\ast})$ is the minimizer in (\ref{s3:max_min_solb}). Provided a condition on $H$ holds, which we specify shortly, the extrema in (\ref{s3:max_min_sol}) are unique, which we henceforth assume.
The functions $u_{1}(\bsx)$ and $u_{2}(\bsy)$ are referred to as a $G$-convex $H$-concave solution pair; see \cite{Romijn2021}. We prove that (\ref{s3:max_min_sola}) together with (\ref{s3:geom_eqb}) implies (\ref{s3:max_min_solb}); the implication in the reverse order proceeds in a similar way. Assume that the expression for $u_{1}(\bsx)$ in (\ref{s3:max_min_sola}) holds, then
\[
  \forall \bsx \in \mcX\, \forall \bsy \in \mcY \quad u_{1}(\bsx) \ge G( \bsx,\bsy,u_{2}(\bsy) ).
\]
Applying the inverse $H( \bsx,\bsy,\cdot ) = G^{-1}( \bsx,\bsy,\cdot )$ and using that $H_{w}( \bsx,\bsy,\cdot ) > 0$, we obtain
\[
  \forall \bsx \in \mcX\, \forall \bsy \in \mcY \quad H( \bsx,\bsy,u_{1}(\bsx) ) \ge u_{2}(\bsy).
\]
Since this inequality holds for all $\bsx \in \mcX$, we can take the minimum and find
\[
  \forall \bsy \in \mcY \quad u_{2}(\bsy) \le \min_{\bsx \in \mcX} H( \bsx,\bsy,u_{1}(\bsx) ).
\]
Let $\bsy = \bsm(\tilde{\bsx}) \in \mcY$ for some $\tilde{\bsx} \in \mcX$.
Then, by virtue of Eq. (\ref{s3:geom_eqb})
\[
  \quad u_{2}(\bsy) = H( \tilde{\bsx},\bsy,u_{1}(\tilde{\bsx}) ) \ge \min_{\bsx \in \mcX} H( \bsx,\bsy,u_{1}(\bsx) ).
\]
Combining both inequalities we arrive at the expression for $u_{2}(\bsy)$ in Eq. (\ref{s3:max_min_solb}) with $\tilde{\bsx} = \bsx^{\ast}$ the unique minimizer. Conversely, given the expression for $u_{2}(\bsy)$ in (\ref{s3:max_min_solb}), we can derive in a similar manner the expression for $u_{1}(\bsx)$ in (\ref{s3:max_min_sola}).

Eq. (\ref{s3:geom_eqb}) also has a $G$-concave $H$-convex (min/max) solution pair. The necessary condition for both solutions is that $(\bsx,\bsm(\bsx))$ is a stationary point of $\widetilde{H}(\bsx,\bsy) = H( \bsx,\bsy,u_{1}(\bsx) )$, i.e.,
\begin{equation} \label{s3:stationary_point1}
  \nabla_{\bsx} \widetilde{H}( \bsx,\bsm(\bsx) ) =
  \nabla_{\bsx} H( \bsx,\bsm(\bsx),u_{1}(\bsx) ) + H_{w}( \bsx,\bsm(\bsx),u_{1}(\bsx) ) \nabla u_{1}(\bsx) = \bzero.
\end{equation}
A sufficient condition for the max/min pair is that $\rmD_{\bsx\bsx} \widetilde{H}(\bsx,\bsm(\bsx))$ be SPD for all $\bsx \in \mcX$. Alternatively, for the min/max pair, the Hessian matrix $\rmD_{\bsx\bsx} \widetilde{H}(\bsx,\bsm(\bsx))$ should be SND for all $\bsx \in \mcX$. In both cases the solution pair is unique.

Equation (\ref{s3:stationary_point1}) implicitly defines the optical map $\bsy = \bsm(\bsx)$ provided the mixed derivative matrix $\rmD_{\bsx\bsy} \widetilde{H}(\bsx,\bsy)$ is regular in $(\bsx,\bsm(\bsx))$ for all $\bsx \in \mcX$, but, the actual computation of the optical map from (\ref{s3:stationary_point1}) is virtually impossible for most optical systems. So, we differentiate the zero-gradient condition (\ref{s3:stationary_point1}) and recover the matrix-Jacobi equation in Eq. (\ref{s3:matrix_eq}), however, with matrices $\bsC$ and $\bsP$ given by
\begin{equation} \label{s3:CP_matrices}
  \bsC(\bsx,\bsy,u_{1}(\bsx)) = \rmD_{\bsx \bsy} \widetilde{H}(\bsx,\bsy),
  \quad \bsP(\bsx,\bsy,u_{1}(\bsx)) = -\rmD_{\bsx\bsx} \widetilde{H}(\bsx,\bsy).
\end{equation}
Notice, there is one difficulty, i.e., the function $\widetilde{H}( \bsx,\bsy )$ depends on $u_{1}(\bsx)$, and consequently, also the matrices $\bsC$ and $\bsP$ do. Therefore, the computation of the optical map and the function $u_{1}(\bsx)$ can no longer be decoupled; see Section \ref{s6:LS} for more details.
Finally, also in this case the transport boundary condition (\ref{s3:transport_bc}) applies.


\section{Conservation of luminous flux}
\label{s4:EC}

In the previous section we presented equations describing the geometry of an optical system, more specifically, Eqs. (\ref{s3:geom_eq}) for the shape/location of the optical surface(s), the zero-gradient conditions (\ref{s3:stationary_point}) and (\ref{s3:stationary_point1}), and the matrix-Jacobi equation (\ref{s3:matrix_eq}) for the optical map. To close the model, we have to formulate the conservation law of luminous flux.

\begin{figure}[!ht]
  \centering
\begin{subfigure}[b]{0.48\textwidth}
    \includegraphics[width = 1.0\textwidth,trim={0cm 0cm 0cm 0cm},clip ]{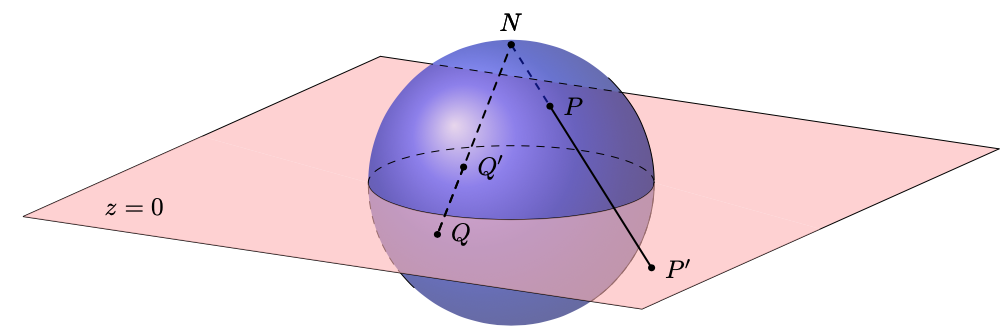}
    \caption{ From the north pole $\rmN$.}
    \label{Fig:projection_north}
\end{subfigure}
\hfill
\begin{subfigure}[b]{0.48\textwidth}
    \includegraphics[width = 1.0\textwidth,trim={0cm 0cm 0cm 0cm},clip ]{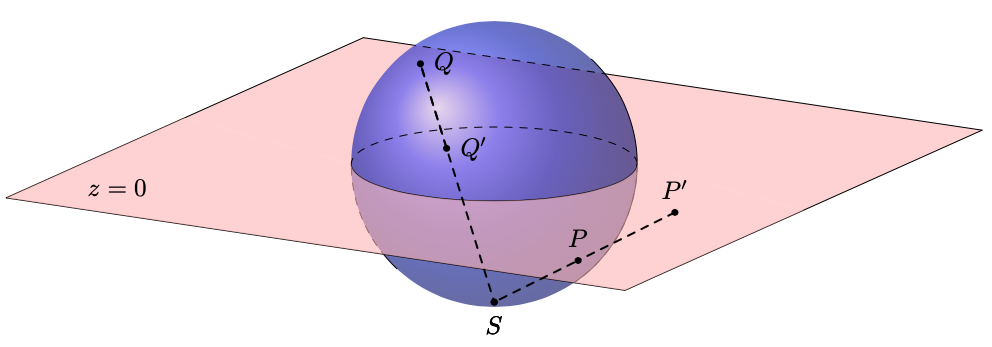}
    \caption{ From the south pole $\rmS$.}
    \label{Fig:projection_south}
\end{subfigure}
\caption{ Stereographic projections of the unit sphere $\rmS^{2}$.}
\label{Fig:projections}
\end{figure}

We first introduce stereographic projections of a unit vector $\hat{\bsv} \in \rmS^{2}$, needed in some of the flux balances. A unit vector $\hat{\bsv}$ can be represented by a point $\rmP$ or $\rmQ$ on the unit sphere; see Figure \ref{Fig:projections}. There are two stereographic projections of $\hat{\bsv}$, i.e., the projection from the north pole and from the south pole. To compute the stereographic projection from the north pole $\rmN$, we have to compute the intersection of the line through $\rmN$ and $\rmP$, given by
\[
  \bsr(\lambda) = \begin{pmatrix} 0 \\ 0 \\ 1 \end{pmatrix} + \lambda \begin{pmatrix} v_{1} \\ v_{2} \\ v_{3}-1 \end{pmatrix},
\]
with the equator plane $z = 0$. This way we obtain the stereographic projection $\bsz = \bfS_{\rmN}(\hat{\bsv})$ given by
\begin{subequations} \label{s4:projection_north}
\begin{equation}
  \bfS_{\rmN}(\hat{\bsv}) = \frac{1}{1-v_{3}} \begin{pmatrix} v_{1} \\ v_{2} \end{pmatrix},
\end{equation}
which is defined for $v_{3} \ne 1$. The latter condition means that $\rmP$ and $\rmN$ should not coincide.
Using the relations $|\hat{\bsv}|^{2} = 1$ and $v_{3} \ne 1$, we can compute the inverse
$\hat{\bsv} = \bfS_{\rmN}^{-1}(\bsz)$, and find
\begin{equation}
  \bfS_{\rmN}^{-1}(\bsz) = \frac{1}{1 + |\bsz|^{2}} \begin{pmatrix} 2 z_{1} \\ 2 z_{2} \\ -1 + |\bsz|^{2} \end{pmatrix}.
\end{equation}
\end{subequations}
In a similar manner we can determine the stereographic projection from the south pole and its inverse. These are given by
\begin{equation} \label{s4:projection_south}
   \bsz = \bfS_{\rmS}(\hat{\bsv}) = \frac{1}{1+v_{3}} \begin{pmatrix} v_{1} \\ v_{2} \end{pmatrix}, \quad
   \hat{\bsv} = \bfS_{\rmS}^{-1}(\bsz) = \frac{1}{1 + |\bsz|^{2}} \begin{pmatrix} 2 z_{1} \\ 2 z_{2} \\ 1 - |\bsz|^{2} \end{pmatrix},
\end{equation}
and are defined for $v_{3} \ne -1$. Both projections are parameterizations of $\rmS^{2}$, and in either case an area element $\rmd S(\bsz)$ on $S^{2}$ generated by $\hat{\bsv}$ is given by
\begin{equation} \label{s4:area_element}
 \rmd S(\bsz) = \Big| \pdfds{\hat{\bsv}}{z_{1}} \crossproduct \pdfds{\hat{\bsv}}{z_{2}} \Big|\, \rmd \bsz =
  \frac{ 4 }{\big( 1 + |\bsz|^{2} \big)^{2}}\, \rmd \bsz =: J(\hat{\bsv}(\bsz))\, \rmd \bsz.
\end{equation}
The choice for the stereographic projection depends on the direction of $\hat{\bsv}$. If $v_{3} > 0$, a suitable choice is the projection from the south pole, since then the domain for $\bsz$ is included in the interior of the disc $|\bsz| \le 1$. The projection from the north pole gives a domain outside this disc; see Figure \ref{Fig:projections}. Likewise, if $v_{3} < 0$, the projection from the north pole is the proper choice.



Assuming there is no energy lost, the most generic form of the luminous flux balance reads
\begin{equation} \label{s4:balance}
  \int_{\mcA} f(\bsx)\, \rmd S(\bsx) = \int_{\bsm(\mcA)} g(\bsy)\, \rmd S(\bsy),
\end{equation}
for every subset $\mcA \subset \mcX$ and image set $\bsm(\mcA) \subset \mcY$, where $f(\bsx)$ and $g(\bsy)$ are the flux densities, to be specified shortly, at the source and target, respectively. Here, $\rmd S(\bsx)$ denotes an area element on the (curved) surface $\mcS$ describing the source, and parameterized by $\bsx$. Analogously, $\rmd S(\bsy)$ is an area element on the target surface $\mcT$, parameterized by $\bsy$. Equation (\ref{s4:balance}) should hold for any subsets $\mcA \subset \mcX$ and $\bsm(\mcA) \subset \mcY$, so also for $\mcA = \mcX$ and $\bsm(\mcA) = \mcY$, giving the global luminous flux balance. This implies that an inverse design problem can only have a solution if $f$ and $g$ satisfy this global flux balance.

The precise form of the flux balance depends on the source and target. For the source we consider two options, first, a planar source located in $z = 0$ and emitting a parallel beam of light with $\hat{\bss} = \hat{\bse}_{z}$, and second, a point source located in the origin and emitting a conical beam of light with $\hat{\bss} = \hat{\bse}_{r}$. Both are zero-\'{e}tendue (English: zero-extent) sources, meaning that the emitted beam has either spatial or angular extent, but in the embedding four-dimensional phase space it has zero volume; see \cite{Chaves2016}. For the planar source $\mcS = \{ \bsr(\bsx) = (\bsx, 0) | \bsx \in \mcX \subset \dR^{2} \}$, $\bsx = \bsq_{\rms}$ are spatial coordinates (Cartesian/polar), $\rmd S(\bsx) = \rmd \bsx$, the area element in the source plane, and $f(\bsx) = M(\bsr(\bsx))$, the emittance, i.e., the luminous flux per unit area emitted by the source. The point source has no spatial extent, but emits light rays in an angular domain, specified by $\hat{\bss} \in \mcS \subset \rmS^{2}$. As parameterization we choose a stereographic projection, either from the north or south pole, the area element $\rmd S(\bsx) = J(\hat{\bss}(\bsx)) \rmd \bsx$, and $f(\bsx) = I_{\rms}( \hat{\bss}(\bsx) )$, the intensity of the source, i.e., the emitted luminous flux per unit solid angle. We can express both flux densities in the form $f = \tilde{f} \circ \bfP_{\rms}^{-1}$, where $\tilde{f} = M$ and $\bfP_{\rms}: (\bsx,0) \mapsto \bsx$, the trivial projection from $\mcS$ to $\mcX$, for the planar source and $\tilde{f} = I_{\rms}$ and $\bfP_{\rms} = \bfS_{\rmN}$ or $\bfP_{\rms} = \bfS_{\rmS}$ for the point source.

For the target we consider the following cases, first a near-field target located on a curved surface
$\mcT = \{ \bsr(\bsy) = (\bsy, v(\bsy)) |\, \bsy \in \mcY \subset \dR^{2} \}$, and second, a far-field target.
For the near-field target $\bsy = \bsq_{\rmt}$ are spatial coordinates in a reference plane $z = z_{\rmt}$, the area element
\[
  \rmd S(\bsy) = \Big| \pdfds{\bsr}{y_{1}} \crossproduct \pdfds{\bsr}{y_{2}} \Big| \rmd \bsy = \sqrt{ |\nabla v(\bsy)|^{2} + 1 }\, \rmd \bsy,
\]
and $g(\bsy) = E(\bsr(\bsy))$, the illuminance, i.e, the luminous flux per unit area incident on the target surface. The domain of a far-field target is an angular domain, determined by the transmitted rays with $\hat{\bst} \in \mcT \subset \rmS^{2}$. Therefore we choose for $\bsy$ one of the two stereographic projections for which
$\rmd S(\bsy) = J( \hat{\bst}(\bsy) ) \rmd \bsy$. However, $g(\bsy) = I_{\rmt}(\hat{\bsr}(\bsy))$, the intensity in the target domain as a function of $\hat{\bsr}$, the unit direction vector directed straight from source to target domain, skipping all optical surfaces, rather than a function of $\hat{\bst}$. In the far-field approximation the distance from source to target is large compared to the size of the optical system, specifically $|\bsr - d \hat{\bst}| / |\bsr| \ll 1$, where $d$ is the distance from optical surface to target, measured along a reflected ray; see, e.g., \cite[pp. 59 - 60]{Romijn2021}. The optical system is considered a point contracted at the origin so we simply take $g(\bsy) = I_{\rmt}(\hat{\bst}(\bsy))$. Also for the target both flux densities can be unified in the expression $g = \tilde{g} \circ \bfP_{\rmt}^{-1}$, where for the near field $\tilde{g} = E$ and $\bfP_{\rmt}: (\bsy, v(\bsy)) \mapsto \bsy$ is the orthogonal projection from $\mcT$ to $\mcY$, and where for the far-field target $\tilde{g} = I_{\rmt}$ and  either $\bfP_{\rmt} = \bfS_{\rmN}$ or $\bfP_{\rmt} = \bfS_{\rmS}$.


\section{Reflector and lens equations: a hierarchy of mathematical models}
\label{s5:hierarchy}

We present a hierarchy of mathematical models for optical systems with zero-\'{e}tendue source, based on three example systems, i.e., parallel-to-far-field reflector, point-to-far-field lens and parallel-to-near-field reflector. In this section, we combine the geometrical equations with the conservation law for luminous flux.

\noindent
\subsection{Parallel-to-far-field reflector}
\label{s5.1:Pa2FFR}

\begin{figure}[!b]
\begin{center}
  \includegraphics[width=0.75\textwidth]{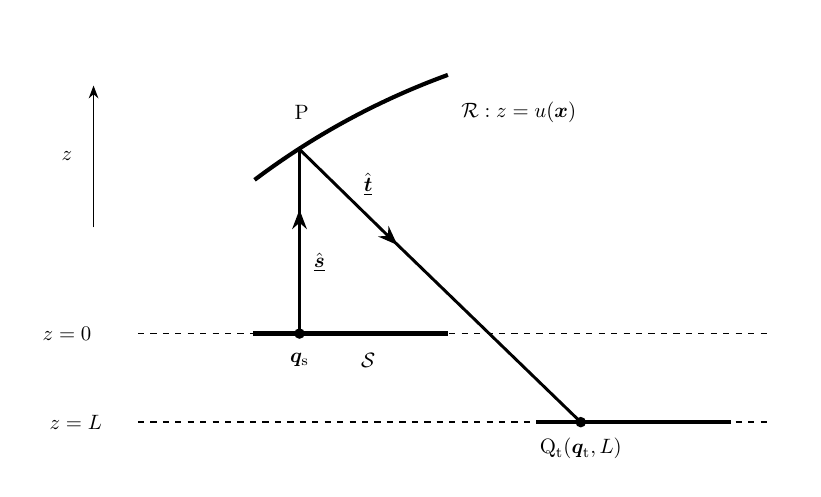}
\end{center}
\caption{ Sketch of a parallel-to-far-field reflector. }
\label{Fig:Pa2FFR}
\end{figure}

\noindent
We consider a light source located in the plane $z = 0$ emitting a parallel beam of light for which $\hat{\bss} = \hat{\bse}_{z}$. Light rays strike the reflector $\mcR$ defined by $z = u(\bsx)$, $\bsx \in \mcX \subset \dR^{2}$, and are reflected off in the direction $\hat{\bst}\in \mcT \subset \rmS^{2}$.
The reflected rays intersect an auxiliary target plane $z = L$; see Figure \ref{Fig:Pa2FFR}. To derive a geometrical equation for the reflector surface, we determine from Eq. (\ref{s2:W_char}) the mixed characteristic of the first kind $W( \bsq_{\rms},\bsp_{\rmt})$ for an arbitrary ray with $\bsq_{\rms} = \bsx$ and $\bsp_{\rmt} = ( t_{1}\; t_{2})^{\rm T}$. The point characteristic $V( \bsq_{\rms},\bsq_{\rmt} )$ involved is given by
\begin{equation} \label{s5:V_char}
  V( \bsq_{\rms},\bsq_{\rmt} ) = [ \rmQ_{\rms},\rmQ_{\rmt} ] = u( \bsx ) + d, \quad
  d = \sqrt{ |\bsq_{\rmt} - \bsx|^{2} + (L-u(\bsx))^{2} },
\end{equation}
where $d = d(\rmP,\rmQ_{\rmt})$ is the distance between the points $\rmP(\bsx,u(\bsx))$ and $\rmQ_{\rmt}(\bsq_{\rmt},L)$, which are the points where the reflected ray hits reflector and target plane, respectively, loosely referred to as intersection points. For the direction vector $\hat{\bst}$ of the reflected ray we have
\begin{equation} \label{s5:t_vector}
  \bsp_{\rmt} = \begin{pmatrix} t_{1} \\ t_{2} \end{pmatrix} = \frac{1}{d} ( \bsq_{\rmt} - \bsx ), \quad t_{3} = \frac{1}{d} ( L - u(\bsx) ).
\end{equation}
Moreover, $\partial W / \partial \bsq_{\rms} = -\bsp_{\rms} = \bzero$, since $\hat{\bss}$ is perpendicular to the source plane, implying that $W = W(\bsp_{\rmt})$. Elaborating the expression for $W$, using the relations in Eqs. (\ref{s5:V_char}) and (\ref{s5:t_vector}) we find
\begin{equation} \label{s5:W_char}
\begin{split}
  W( \bsp_{\rmt} ) &= u(\bsx) + d - \bsq_{\rmt} \dotproduct \bsp_{\rmt}\\
  &= u(\bsx) + \frac{1}{d}\big( |\bsq_{\rmt}-\bsx|^{2} + (L-u(\bsx))^{2} - \bsq_{\rmt} \dotproduct (\bsq_{\rmt}-\bsx) \big)\\
  &= u(\bsx) + \frac{1}{d} \big( -\bsx \dotproduct (\bsq_{\rmt} - \bsx) + d t_{3}(L-u(\bsx)) \big)\\
  &= u(\bsx) (1-t_{3}) - \bsx \dotproduct \bsp_{\rmt} + L t_{3}.
\end{split}
\end{equation}


To separate the variables $\bsx$ (source) and $\bsp_{\rmt}$ (target), we bring all terms that solely depend on $\hat{\bst}$ to the left-hand side, and we obtain
\begin{equation} \label{s5:separation}
  \frac{ W(\bsp_{\rmt}) - L t_{3} }{ 1-t_{3} } = u(\bsx) - \frac{\bsx \dotproduct \bsp_{\rmt}}{1-t_{3}}.
\end{equation}
Note that $t_{3} \ne 1$ since rays cannot pass straight through the reflector.
Using the expressions for $d$ and $t_{3}$ we can show that
\[
  \pdfds{W}{L} = \pdfds{d}{L} - \pdfds{\bsq_{\rmt}}{L} \dotproduct \bsp_{\rmt}, \quad \pdfds{d}{L} = \bsp_{\rmt} \dotproduct \pdfds{\bsq_{\rmt}}{L} + t_{3}
\]
from which we readily conclude that the left-hand side in Eq. (\ref{s5:separation}) is independent of $L$, which makes sense since a far-field target domain should not depend on the choice of a particular plane $z = L$.
As stated in Section \ref{s4:EC} the proper choice to parameterize the target $\mcT$ is the stereographic projection from the north pole, thus $\bsy = \bsp_{\rmt}/(1-t_{3})$ and $\mcY = \bfS_{\rmN}(\mcT)$; cf. Eq. (\ref{s4:projection_north}).
Substituting the expression for $\bsy$ in Eq. (\ref{s5:separation}), we arrive at the geometrical equation
\begin{equation} \label{s5:OT_formulation}
  u_{1}(\bsx) + u_{2}(\bsy) = c(\bsx,\bsy), \quad c(\bsx,\bsy) = \bsx \dotproduct \bsy,
\end{equation}
where $u_{1}(\bsx) = u(\bsx)$ and $u_{2}(\bsy) = ( L t_{3} - W(\bsp_{\rmt}) )/( 1-t_{3} )$.

Referring to Section \ref{s3:geometry}, a possible solution of Eq. (\ref{s5:OT_formulation}) is the max/max solution in (\ref{s3:max_max_sol}), which is a conjugate pair of Legendre-Fenchel tranforms since $c(\bsx,\bsy) = \bsx \dotproduct \bsy$. The necessary condition in (\ref{s3:stationary_point}) and the matrix equation for the optical map in (\ref{s3:matrix_eq}) reduce to
\begin{subequations} \label{s5:geom_eq}
\begin{align}
  &\bsy - \nabla u_{1}(\bsx) = \bzero,\\
  &\rmD \bsm = \bsP(\bsx), \quad \bsP(\bsx) = \rmD^{2} u_{1}(\bsx). \label{s5:geom_eqb}
\end{align}
\end{subequations}
Recall that equation (\ref{s5:geom_eqb}) has to be supplemented with the transport boundary condition (\ref{s3:transport_bc}).

To close the model, we elaborate the flux balance in Eq. (\ref{s4:balance}). Substituting $f(\bsx) = M(\bsr(\bsx))$, $\rmd S(\bsx) = \rmd \bsx$, $g(\bsy) = I_{\rmt}(\hat{\bst}(\bsy))$ and $\rmd S(\bsy) = J(\hat{\bst}(\bsy)) \rmd \bsy$, we obtain
\begin{equation} \label{s5:balance1}
  \int_{\mcA} M(\bsr(\bsx))\, \rmd \bsx = \int_{\bsm(\mcA)} I_{\rmt}( \hat{\bst}(\bsy) ) J( \hat{\bst}(\bsy) )\, \rmd \bsy,
\end{equation}
for arbitrary $\mcA \subset \mcX$ and $\bsm(\mcA) \subset \mcY$. Substituting $\bsy = \bsm(\bsx)$ and the expression for $J( \hat{\bst}(\bsy) )$, replacing $\hat{\bsv}(\bsz)$ by $\hat{\bst}(\bsy)$ in Eq. (\ref{s4:area_element}), we can rewrite the integral in the right-hand side, and find
\begin{equation} \label{s5:balance2}
  \int_{\mcA} M(\bsr(\bsx))\, \rmd \bsx = \int_{\mcA} I_{\rmt}( \hat{\bst}(\bsm(\bsx)) ) \frac{4}{\big( 1 + |\bsm(\bsx)|^{2} \big)^{2} } \det(\rmD \bsm)\, \rmd \bsx,
\end{equation}
where we used that $\det(\rmD \bsm) > 0$, which is correct because $\rmD \bsm$ is either SPD or SND.
Since Eq. (\ref{s5:balance2}) should hold for arbitrary $\mcA \subset \mcX$, we obtain the differential form
\begin{equation} \label{s5:Jacobian_eq}
  \det( \rmD \bsm ) = \tfrac{1}{4} \big( 1 + |\bsm(\bsx)|^{2} \big)^{2} \frac{ M(\bsr(\bsx)) }{ I_{\rmt}(\hat{\bst}(\bsm(\bsx))) } =: F( \bsx,\bsm(\bsx) ),
\end{equation}
referred to as the Jacobian equation. Substituting $\bsm = \nabla u_{1}$, this equation reduces to the standard MA equation $\det( \rmD^{2} u_{1} ) = F( \bsx,\nabla u_{1}(\bsx) )$; see, e.g., \cite{Gutierrez2016}.

\noindent
\subsection{Point-to-far-field lens}
\label{s5.2:Po2FFL}

\begin{figure}[!t]
\begin{center}
  \includegraphics[width=0.6\textwidth]{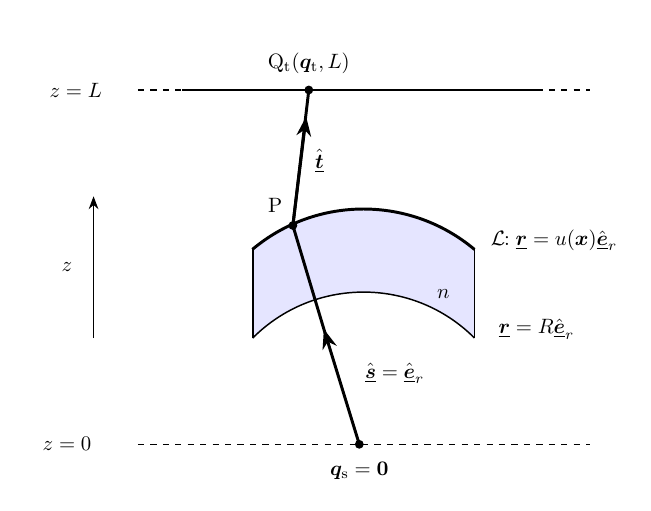}
\end{center}
\caption{ Sketch of a point-to-far-field lens. }
\label{Fig:Po2FFL_sketch}
\end{figure}

\noindent
We consider a point source located in the origin $\rmO_{\rms}$ (position vector given by $\bsq_{\rms} = \bzero$ and $z_{\rms} = 0$),  emitting upwards a conical beam of light with direction vector $\hat{\bss} = \hat{\bse}_{r}$. Light rays hit a lens of refractive index $n$; see Figure \ref{Fig:Po2FFL_sketch}. The first (entrance) surface is spherical with centre $\rmO_{\rms}$ and radius $R$, hence rays are not refracted there, however, the second (exit) surface is freeform and given by the parameterization $\bsr(\hat{\bss}) = u(\hat{\bss}) \hat{\bss}$. Rays are refracted at the exit surface and arrive at an auxiliary target screen $z = L > 0$. To derive a geometrical equation for the freeform surface, we evaluate the angular characteristic
$T( \bsp_{\rms},\bsp_{\rmt} )$ defined in (\ref{s2:T_char}) for an arbitrary ray connecting source with target plane. The point characteristic $V( \bsq_{\rms},\bsq_{\rmt} )$ needed to evaluate $T(\bsp_{\rms},\bsp_{\rmt})$ is given by
\begin{equation} \label{s5:V_char1}
  V( \bsq_{\rms},\bsq_{\rmt} ) = [ \rmO_{\rms},\rmQ_{\rmt} ] =  R + n(u(\hat{\bss}) - R) + d, \quad
  d = \sqrt{ |\bsq_{\rmt} - u(\hat{\bss}) \bsp_{\rms}|^{2} +
  (L-u(\hat{\bss}) s_{3})^{2} },
\end{equation}
where $d = d(\rmP,\rmQ_{\rmt})$ is the distance between the points $\rmP( u(\hat{\bss}) \bsp_{\rms}, u(\hat{\bss}) s_{3})$ and $\rmQ_{\rmt}( \bsq_{\rmt},L )$, which are the intersection points of the refracted ray with exit surface and target screen, respectively. The direction vector $\hat{\bst}$ of the refracted ray is given by
\begin{equation} \label{s5:t_vector1}
  \bsp_{\rmt} = \begin{pmatrix} t_{1} \\ t_{2} \end{pmatrix} = \frac{1}{d} ( \bsq_{\rmt} - u(\hat{\bss}) \bsp_{\rms} ), \quad t_{3} = \frac{1}{d} ( L - u(\hat{\bss}) s_{3}).
\end{equation}
Note that $\partial T / \partial \bsp_{\rms} = \bsq_{\rms} = \bzero$, implying that
$T = T( \bsp_{\rmt} )$. Evaluating the expression for $T( \bsp_{\rmt} )$ in Eq. (\ref{s2:T_char}) using the relations in Eqs. (\ref{s5:V_char1}) and (\ref{s5:t_vector1}), we obtain
\begin{equation} \label{s5:T_char}
\begin{split}
  T( \bsp_{\rmt} ) &= R + n( u(\hat{\bss}) - R ) + d - \bsq_{\rmt} \dotproduct \bsp_{\rmt}\\
  &=  (1-n) R + n u(\hat{\bss}) + \frac{1}{d}\big( |\bsq_{\rmt}-u(\hat{\bss}) \bsp_{\rms}|^{2} + (L-u(\hat{\bss}) s_{3})^{2} - \bsq_{\rmt} \dotproduct (\bsq_{\rmt}-u(\hat{\bss}) \bsp_{\rms}) \big)\\
  &= (1-n) R + n u(\hat{\bss}) + \frac{1}{d} \big( -u(\hat{\bss}) \bsp_{\rms} \dotproduct (\bsq_{\rmt} - u(\hat{\bss}) \bsp_{\rms}) + d t_{3}(L-u(\hat{\bss}) s_{3}) \big)\\
  &= (1-n) R + n u(\hat{\bss}) - u(\hat{\bss})\bsp_{\rms} \dotproduct \bsp_{\rmt} + t_{3} ( L-u(\hat{\bss}) s_{3})\\
  &= (1-n) R + ( n - \hat{\bss} \dotproduct \hat{\bst} ) u(\hat{\bss}) + L t_{3}.
\end{split}
\end{equation}
To derive the last expression for $T$ in (\ref{s5:T_char}) we substituted the relation $\hat{\bss} \dotproduct \hat{\bst} = \bsp_{\rms} \dotproduct \bsp_{\rmt} + s_{3} t_{3}$.
Next, we move all terms that solely depend on $\hat{\bst}$ and the constant $(1-n) R$ to the left-hand side and find
\begin{equation} \label{s5:separation1}
  T(\bsp_{\rmt}) - L t_{3} + (n-1) R = ( n - \hat{\bss} \dotproduct \hat{\bst} ) u(\hat{\bss}).
\end{equation}
Differentiating the first expression in (\ref{s5:T_char}) with respect to $L$ and combining the expression for $d$ with (\ref{s5:t_vector1}) we obtain
\[
  \pdfds{T}{L} = \pdfds{d}{L} - \pdfds{\bsq_{\rmt}}{L} \dotproduct \bsp_{\rmt}, \quad\pdfds{d}{L} = \bsp_{\rmt} \dotproduct \pdfds{\bsq_{\rmt}}{L} + t_{3},
\]
from which we readily see that the left-hand side in (\ref{s5:separation1}) is independent of $L$, as anticipated.

To separate source and target coordinates, analogous to Eq. (\ref{s5:OT_formulation}), we have to take the logarithm of the left- and right-hand sides of equation (\ref{s5:separation1}). Note that $n - \hat{\bss} \dotproduct \hat{\bst} > 0$ and $u(\hat{\bss}) > 0$, consequently, the left-hand side is positive as well, and we can take the logarithms. In addition, we substitute the stereographic projections (from the south pole) $\bsx$ of $\hat{\bss}$ and $\bsy$ of $\hat{\bst}$, since both $\hat{\bss}$ and $\hat{\bst}$ are directed upwards; cf. Eq. (\ref{s4:projection_south}). This way, we obtain
\begin{subequations}
\begin{equation} \label{s5:OT_formulation1}
  u_{1}(\bsx) + u_{2}(\bsy) = c(\bsx,\bsy),
\end{equation}
where the variables $u_{1}(\bsx)$ and $u_{2}(\bsy)$ and the cost function $c( \bsx,\bsy )$ are defined by
\begin{align}
  u_{1}(\bsx) &= \log u(\hat{\bss}(\bsx)), \label{s3:opt_transport_a}\\
  u_{2}(\bsy) &= -\log\big( T(\bsp_{\rmt}(\bsy)) - L t_{3}(\bsy) + (n-1) R \big),\\
  c( \bsx,\bsy ) &= -\log\big( n - \hat{\bss}(\bsx) \dotproduct \hat{\bst}(\bsy) \big).
\end{align}
\end{subequations}
Formulated in terms of stereographic coordinates the cost function reads
\[
  c( \bsx,\bsy) = -\log\Big( n - 1 + \frac{ 2 |\bsx-\bsy|^{2} }{( 1 + |\bsx|^{2} )( 1 + |\bsy|^{2} )} \Big),
\]
which is no longer quadratic. A possible solution of Eq. (\ref{s5:OT_formulation1}) is the $c$-convex pair in (\ref{s3:max_max_sol}), for which the necessary condition (\ref{s3:stationary_point}) holds. For the optical map, the matrix equation (\ref{s3:matrix_eq}) accompanied with the transport boundary condition (\ref{s3:transport_bc}) is given.

Referring to Section \ref{s4:EC}, in the far-field approximation the luminous flux balance reads
\begin{equation} \label{s5:flux_balance31}
  \int_{\mcA} I_{\rms}(\hat{\bss}(\bsx)) J(\hat{\bss}(\bsx))\, \rmd \bsx = \int_{\bsm(\mcA)} I_{\rmt}(\hat{\bst}(\bsy)) J(\hat{\bst}(\bsy))\, \rmd \bsy,
\end{equation}
for arbitrary $\mcA \subset \mcX$ and image set $\bsm(\mcA) \subset \mcY$. To derive the differential form, we substitute the expressions for $J(\hat{\bss}(\bsx))$ and $J(\hat{\bst}(\bsy))$ according to (\ref{s4:area_element}) and subsequently substitute $\bsy = \bsm(\bsx)$. Assuming $\det( \rmD \bsm ) > 0$, we obtain
\begin{equation}
 \det( \rmD \bsm ) = \left( \frac{ 1 + |\bsm(\bsx)|^{2} }{ 1 + |\bsx|^{2} } \right)^{2} \frac{ I_{\rms}(\bsx) }{ I_{\rmt}(\bsm(\bsx)) } =: F(\bsx,\bsm(\bsx)).
\end{equation}
Combining this equation with the matrix equation in (\ref{s3:matrix_eq}) we obtain the GMA equation $\det( \rmD^{2} u_{1} - \rmD_{\bsx\bsx} c ) = \det( C(\bsx,\bsm(\bsx)) F( \bsx,\bsm(\bsx) )$.

\noindent
\subsection{Parallel-to-near-field reflector}
\label{s5.3:Pa2NFR}

\begin{figure}[!b]
\begin{center}
  \includegraphics[width=0.75\textwidth]{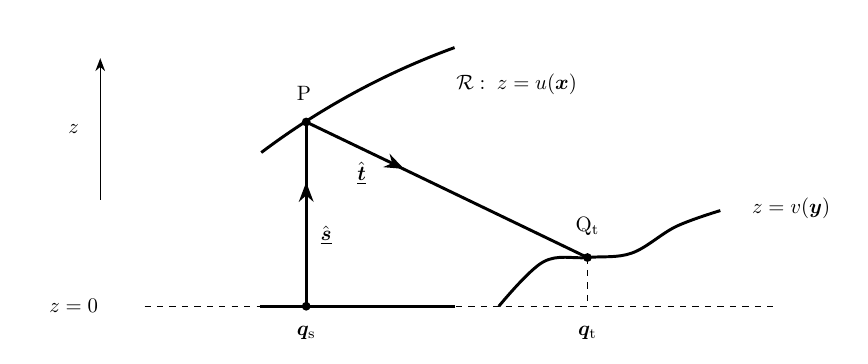}
\end{center}
\caption{ Sketch of a parallel-to-near-field reflector. }
\label{Fig:Pa2NFR}
\end{figure}

The last example concerns a planar source, located in $z = 0$, emitting a parallel beam of light rays in the direction $\hat{\bss} = \hat{\bse}_{z}$, consequently $\bsp_{\rms} = \bzero$, which hits a reflector $\mcR$ given by $z = u(\bsx)$, with $\bsx = \bsq_{\rms}$ spatial coordinates, and are reflected off in the direction $\hat{\bst}$ and lands at a target surface given by $z = v(\bsy)$,  with $\bsy = \bsq_{\rmt}$ spatial coordinates in the reference plane $z = z_{\rmt} = 0$. Thus, source and target planes coincide; see Figure \ref{Fig:Pa2NFR}. We evaluate Hamilton's point characteristic and find
\begin{equation} \label{s5:V_char2}
\begin{split}
  V( \bsq_{\rms},\bsq_{\rmt} ) &= [ \rmQ_{\rms},\rmQ_{\rmt} ]\\
  &= u(\bsx) + d, \quad d = \sqrt{ |\bsy-\bsx|^{2} + (v(\bsy) - u(\bsx))^{2} },
\end{split}
\end{equation}
where $d = d( \rmP,\rmQ_{\rmt} )$ is the distance between the intersection points $\rmP( \bsx,u(\bsx) )$ and $\rmQ_{\rmt}( \bsy,v(\bsy) )$ of the reflected ray with the reflector and the target surface, respectively. Since $\partial V / \partial \bsq_{\rms} = - \bsp_{\rms} = \bzero$, we find that $V = V( \bsq_{\rmt} )$. In this case it is no longer possible to separate the source and target coordinates $\bsx$ and $\bsy$ like in Eqs. (\ref{s5:OT_formulation}) or (\ref{s5:OT_formulation1}). Instead, we write
\begin{equation} \label{s5:H_func}
  u_{2}(\bsy) = H( \bsx,\bsy,u_{1}(\bsx) ),\quad
  H( \bsx,\bsy,w ) = w + \sqrt{ |\bsy-\bsx|^{2} + (v(\bsy)-w)^{2} },
\end{equation}
where $u_{1}(\bsx) = u(\bsx)$ and $u_{2}(\bsy) = V(\bsq_{\rmt})$.
Straightforward differentiation gives $H_{w}(\bsx,\bsy,\cdot) > 0$, implying that for fixed $\bsx, \bsy$ the inverse $G(\bsx,\bsy,\cdot) = H^{-1}(\bsx,\bsy,\cdot)$ exists.
Therefore, we can explicitly determine $u_{1}(\bsx)$ from (\ref{s5:H_func}), and we obtain
\begin{equation} \label{s5:G_func}
  u_{1}(\bsx) = G( \bsx,\bsy,u_{2}(\bsy) ),\quad
  G( \bsx,\bsy,w ) = \tfrac{1}{2} \Big( w + v(\bsy) - \frac{ |\bsy-\bsx|^{2} }{ w-v(\bsy) } \Big).
\end{equation}
A possible solution of Eqs. (\ref{s5:H_func}) and (\ref{s5:G_func}) is the  $G$-convex $H$-concave pair in (\ref{s3:max_min_sol}) for which the necessary condition (\ref{s3:stationary_point1}) holds. The matrix equation for $\bsm$ is given in (\ref{s3:matrix_eq}) with matrices $\bsC$ and $\bsP$ defined in (\ref{s3:CP_matrices}). Recall that these matrices explicitly depend on $u_{1}(\bsx)$. Obviously, also the transport boundary condition (\ref{s3:transport_bc}) holds.

Making the proper choices for the flux densities and area elements, the flux balance (\ref{s4:balance}) reduces to
\begin{equation} \label{s5:balance4}
  \int_{\mcA} M(\bsr(\bsx))\, \rmd \bsx = \int_{\bsm(\mcA)} E(\bsr(\bsy)) \sqrt{ |\nabla v(\bsy)|^{2} + 1 }\, \rmd \bsy
\end{equation}
for an arbitrary set $\mcA \subset \mcX$ and image set $\bsm(\mcA) \subset \mcY$, where $M(\bsr(\bsx))$ and $E(\bsr(\bsy))$ denote the emittance and illuminance of source and target, respectively.
Substituting $\bsy = \bsm(\bsx)$, assuming $\det(\rmD \bsm) > 0$, we obtain
\begin{equation} \label{s5:Jacobian_eq1}
  \det(\rmD \bsm) = \frac{1}{\sqrt{ |\nabla v(\bsm(\bsx))|^{2} + 1 } }\frac{ M(\bsr(\bsx)) }{ E(\bsr(\bsm(\bsx))) } =: F( \bsx,\bsm(\bsx)).
\end{equation}
Combining this equation with Eq. (\ref{s3:matrix_eq}) and the matrices $\bsC$ and $\bsP$ defined in Eq. (\ref{s3:CP_matrices}), we obtain the equation $\det( \rmD_{\bsx\bsx} \widetilde{H} ) = \det( \bsC(\bsx,\bsm(\bsx),u_{1}(\bsx) ) F( \bsx,\bsm(\bsx) )$, referred to as the GJ equation.

\noindent
\subsection{Summary of mathematical models}
\label{s5.4:summary}

All mathematical models considered consist of a geometrical equation, a condition for a stationary point, a matrix equation for the optical map coupled to the transport boundary condition and a luminous flux balance
$\det(\rmD \bsm) = F(\bsx,\bsm(\bsx))$. The conditions for the stationary point read
\begin{subequations} \label{s5:gradients}
\begin{alignat}{2}
  &\text{cost function} \quad&& \nabla_{\bsx}c(\bsx,\bsy) - \nabla u_{1}(\bsx) = \bzero, \label{s5:gradientsa}\\
  &\text{generating function} \quad&& \nabla_{\bsx} \widetilde{H}(\bsx,\bsy) = \bzero. \label{s5:gradientsb}
\end{alignat}
\end{subequations}
Substituting the optical map $\bsy = \bsm(\bsx)$ and differentiating with respect to $\bsx$ gives the matrix equation $\bsC \rmD \bsm = \bsP$ where the matrices $\bsC$ and $\bsP$ are given by
\begin{subequations} \label{s6:CP_matrices}
\begin{alignat}{3}
  &\text{quadratic cost function} \quad&& \bsC = \bsI, \quad&& \bsP = \rmD^{2} u_{1},\\
  &\text{general cost function} \quad&& \bsC = \rmD_{\bsx\bsy} c, \quad&& \bsP = \rmD^{2} u_{1} - \rmD_{\bsx\bsx} c,\\
  &\text{generating function} \quad&& \bsC = \rmD_{\bsx\bsy} \widetilde{H}, \quad&& \bsP = -\rmD_{\bsx\bsx} \widetilde{H} \label{s6:CP_matricesc}.
\end{alignat}
\end{subequations}
Combining the matrix equation for $\bsm$ with the flux balance gives
\[
  \det(\bsC) F(\bsx,\bsm(\bsx)) = \det(\bsP),
\]
which we refer to as the luminous flux constraint.
Note that for the cost-function model, the matrices $\bsC$ and $\bsP$ explicitly depend on $\bsx$ and $\bsm(\bsx)$, on the other hand, for the generating-function model, both matrices in addition depend on $u_{1}(\bsx)$.
In \cite{Anthonissen2021} we have given a similar overview of mathematical models of sixteen base optical systems.

In the next section we outline numerical methods to compute the optical map $\bsy = \bsm(\bsx)$ and the auxiliary variable $u_{1}(\bsx)$. The shape/location of the optical surface is then reconstructed from a simple algebraic equation, i.e., $u(\bsx)= u_{1}(\bsx)$ for the parallel-to-far-field and parallel-to-near-field reflectors and $u(\hat{\bss}(\bsx)) = \rme^{u_{1}(\bsx)}$ for the parallel-to-far-field lens.


\section{Iterative least-squares methods}
\label{s6:LS}

In this section we outline iterative least-squares methods for the models presented in the previous section, starting with the base scheme for the cost-function models and adding modifications for the generation-function model. A detailed account of these methods is presented in a series of theses, see \cite{Prins2014,Yadav2018,Romijn2021}, and papers, see \cite{Prins2015,Beltman2018,Yadav2019a,Yadav2019b,Romijn2019,Romijn2020,Romijn2021a,Romijn2021b,Anthonissen2021}.
\vspace{0.1cm}

\noindent
\textbf{Base scheme for cost function.}
To compute the layout of the optical system we apply a two-stage method, i.e., we first compute the optical map $\bsm$ and subsequently the auxiliary variable $u_{1}$ and possibly $u_{2}$, from which we trivially can compute the shape/location of the optical surface(s). We employ a uniform rectangular grid covering the parameter space of the source domain.

We iteratively compute a symmetric matrix $\bsP$ as approximation of the symmetric part of  $\bsC \rmD \bsm$, satisfying $\det(\bsP) = \det(\bsC(\cdot,\bsm)) F(\cdot,\bsm)$, and a vector field $\bsb: \partial \mcX \rightarrow \partial \mcY$, from which we subsequently compute $\bsm$.
Our solution strategy is then to minimize the functionals
\begin{subequations} \label{s6:functionals}
\begin{align}
  J_{\rmI}[\bsm,\bsP] &= \tfrac{1}{2} \int_{\mcX} |\!| \bsC \rmD \bsm - \bsP |\!|_{\rmF}^{2}\, \rmd \bsx, \notag\\
  \qquad &\text{subject to} \quad \det(\bsP) = \det(\bsC(\cdot,\bsm)) F(\cdot,\bsm),\\
  J_{\rmB}[\bsm,\bsb] &= \tfrac{1}{2} \oint_{\partial \mcX} | \bsm-\bsb |_{2}^{2}\, \rmd s,\\
  J[\bsm,\bsP,\bsb] &= \alpha J_{\rmI}[\bsm,\bsP] + (1-\alpha) J_{\rmB}[\bsm,\bsb], \quad (0 < \alpha < 1), \label{s6:functionalsc}
\end{align}
\end{subequations}
where $|\!| \cdot |\!|_{\rmF}$ denotes the Frobenius norm. The minimization of $J_{\rmB}[\bsm,\bsb]$ is to enforce the transport boundary condition (\ref{s3:transport_bc}). Given an initial guess $\bsm^{0}$, the iteration scheme then reads
\begin{subequations} \label{s6:iter}
\begin{align}
  \bsP^{k+1} &= \mathrm{argmin}_{\bsP \in \mcP(\bsm^{k})} J_{\rmI}[\bsm^{k},\bsP],\\
  \bsb^{k+1} &= \mathrm{argmin}_{\bsb \in \mcB} J_{\rmB}[\bsm^{k},\bsb],\\
  \bsm^{k+1} &= \mathrm{argmin}_{\bsm \in \mcM} J[\bsm,\bsP^{k+1},\bsb^{k+1}], \label{s3:iterc}
\end{align}
\end{subequations}
where the corresponding function spaces are given by
\begin{subequations} \label{s6:spaces}
\begin{align}
  \mcP(\bsm) &= \{ \bsP \in  C^{1}(\mcX)^{2 \times 2} \big| \bsP^{\rmT} = \bsP, \det(\bsP) = \det(\bsC(\cdot,\bsm(\cdot))) F(\cdot,\bsm(\cdot))  \},\\
  \mcB &= \{ \bsb \in C(\partial \mcX)^{2} \big| \bsb(\bsx) \in \partial \mcY \},\\
  \mcM &= C^{2}(\mcX)^{2}.
\end{align}
\end{subequations}
In common parlance, $\mcP(\bsm)$ is the space of $2 \times 2$ matrix functions, that are continuously differentiable on $\mcX$, are symmetric and satisfy the luminous flux constraint. $\mcB$ is the space of continuous vector functions, that map the boundary $\partial \mcX$ to the boundary $\partial \mcY$.

The minimization of $J_{\rmI}[\bsm,\bsP]$ to compute $\bsP$ can be performed point-wise and requires the solution of a constrained minimization problem. $\bsP$ has to be either SPD or SND, which can be enforced by a constraint on $\mathrm{tr}(\bsP)$. The minimization of $J_{\rmB}[\bsm,\bsb]$ to compute $\bsb$ is a piecewise projection of $\bsm$ on $\partial \mcX$. For the minimization of $J[\bsm,\bsP,\bsb]$ we  impose the first variation with respect to $\bsm$ to vanish, i.e.,
\begin{equation} \label{s6:first_var}
  \delta J[\bsm,\bsP,\bsb](\bseta) = \lim_{\varepsilon \rightarrow 0} \frac{1}{\varepsilon} \big( J[ \bsm + \varepsilon \bseta, \bsP, \bsb ] - J[ \bsm,\bsP,\bsb ] \big) = 0,
\end{equation}
for arbitrary $\bseta \in C^{2}(\mcX)^{2}$. Evaluating this limit, applying Gauss's theorem and the fundamental lemma of calculus of variations \cite[p. 185]{Courant1953}, we derive the following coupled boundary value problem (BVP) for $\bsm$
\begin{subequations} \label{s6:BVP_m}
\begin{align}
  \nabla \dotproduct( \bsC^{\rmT} \bsC \rmD \bsm ) &= \nabla \dotproduct (\bsC^{\rmT} \bsP ), \quad \bsx \in \mcX,\\
  (1-\alpha) \bsm + \alpha ( \bsC^{\rmT} \bsC \rmD \bsm ) \hat{\bsn} &= (1-\alpha) \bsb + \alpha \bsC^{\rmT} \bsP \hat{\bsn}, \quad \bsx \in \partial \mcX,
\end{align}
\end{subequations}
where $\hat{\bsn}$ is the unit outward normal on $\partial \mcX$. The divergence of a $2 \times 2$ matrix function $\bsA = ( \bsa_{1}\; \bsa_{2})$ is defined as $\nabla \dotproduct \bsA = \begin{pmatrix} \nabla \dotproduct \bsa_{1} \\ \nabla \dotproduct \bsa_{2} \end{pmatrix}$. For discretization of (\ref{s6:BVP_m}) we employ the standard finite volume method \cite[pp. Appendix B]{Romijn2021}.

Next, upon convergence of the iteration (\ref{s6:iter}), we compute $u_{1}$ from (\ref{s5:gradientsa}) by minimizing the functional
\begin{equation}
   I[u_{1}] = \tfrac{1}{2} \int_{\mcX} | \nabla_{\bsx} c(\cdot,\bsm) - \nabla u_{1}|_{2}^{2}\, \rmd \bsx.
\end{equation}
Analogous to the derivation of (\ref{s6:BVP_m}), we set the first variation $\delta I[u_{1}](v) = 0$ for arbitrary $v \in C^{2}(\mcX)$, to obtain the Neumann problem
\begin{subequations} \label{s6:BVP_u1}
\begin{align}
  \nabla^{2} u_{1} &= \nabla \dotproduct \nabla_{\bsx} c(\cdot,\bsm), \quad \bsx \in \mcX,\\
  \nabla u_{1} \dotproduct \hat{\bsn} &= \nabla_{\bsx} c(\cdot,\bsm) \dotproduct \hat{\bsn}, \quad \bsx \in \partial \mcX.
\end{align}
\end{subequations}
We employ standard finite differences to (\ref{s6:BVP_u1}). $u_{1}$ is determined up to an additive constant, which translates in an additive constant in the location of the parallel-to-far-field reflector, and a multiplicative constant for the location of the point-to-far-field lens. To compute a unique solution from (\ref{s6:BVP_u1}) we either fix the distance from source to optical surface along a specific ray, or prescribe an average for $u_{1}$.
\vspace{0.1cm}

\noindent
\textbf{Extension to generating function.}
The optical map $\bsm$ and the variable $u_{1}$ have to be computed simultaneously since the matrices $\bsC$ and $\bsP$ depend on both variables. Like in the previous case, the optical map satisfies the equation $\bsC \rmD \bsm = \bsP$, with the matrices $\bsC$ and $\bsP$ defined in (\ref{s6:CP_matricesc}), and the luminous flux balance $\det(\rmD \bsm) = F$. Consequently, the functional $J_{\rmI}[\bsm,\bsP]$ remains the same, albeit with different matrices $\bsC$ and $\bsP$. The variable $u_{1}$ has to be computed from equation (\ref{s5:gradientsb}), for which we introduce the functional
\begin{equation} \label{s6:funcI}
\begin{split}
  I[u_{1},\bsm] &= \tfrac{1}{2} \int_{\mcX} \big| \nabla_{\bsx} \widetilde{H}(\cdot,\bsm) \big|_{2}^{2}\, \rmd \bsx\\
  &=\tfrac{1}{2} \int_{\mcX} \big| \nabla_{\bsx} H(\cdot,\bsm,u_{1}) + H_{w}(\cdot,\bsm,u_{1}) \nabla u_{1} \big|_{2}^{2}\, \rmd \bsx.
\end{split}
\end{equation}
To minimize the functional in (\ref{s6:funcI}) we require $\delta I[u_{1},\bsm](v) = 0$ for arbitrary $v \in C^{2}(\mcX)$, with $\delta I[u_{1},\bsm](v)$ the first variation of $I$ with respect to $u_{1}$, defined analogously to (\ref{s6:first_var}). This way we can derive the BVP
\begin{subequations} \label{s6:BVP_u11}
\begin{align}
  \nabla \dotproduct \big( H_{w} \nabla_{\bsx} H + H_{w}^{2} \nabla u_{1} \big) &= \tfrac{1}{2} \dfds{ }{w}
  \big| \nabla_{\bsx} H + H_{w} \nabla u_{1} \big|_{2}^{2}, \quad \bsx \in \mcX,\\
  H_{w} \big( \nabla_{\bsx} H + H_{w} \nabla u_{1} \big) \dotproduct \hat{\bsn} &= 0, \quad \bsx \in \partial \mcX.
\end{align}
\end{subequations}
We employ the standard finite volume method for discretization. We can show that also the solution of (\ref{s6:BVP_u11})
is determined up to an additive constant, and compute a unique solution prescribing the average value of $u_{1}$; see \cite[pp. 166 - 167]{Romijn2021} for more details.
Finally, given an initial guess $\bsm^{0}$ and $u_{1}^{0}$, the iteration scheme then reads
\begin{subequations} \label{s6:iter1}
\begin{align}
  \bsP^{k+1} &= \mathrm{argmin}_{\bsP \in \mcP(\bsm^{k})} J_{\rmI}[\bsm^{k},u_{1}^{k},\bsP],\\
  \bsb^{k+1} &= \mathrm{argmin}_{\bsb \in \mcB} J_{\rmB}[\bsm^{k},\bsb],\\
  \bsm^{k+1} &= \mathrm{argmin}_{\bsm \in \mcM} J[\bsm,u_{1}^{k},\bsP^{k+1},\bsb^{k+1}],\\
  u_{1}^{k+1} &= \mathrm{argmin}_{v \in \mcU} I[v,\bsm^{k+1}],
\end{align}
\end{subequations}
with $\mcU = C^{2}(\mcX)$, where we have included $u_{1}^{k}$ in the argument list of $J_{\rmI}$ and $J$ to denote their implicit dependence on $u_{1}$ via the matrix $\bsC$.


\section{Numerical examples}
\label{s7:Examples}

We present two numerical examples, viz. the point-to-far-field lens discussed in Section \ref{s5.2:Po2FFL} and the parallel-to-near-field reflector from Section \ref{s5.3:Pa2NFR}.

\noindent
\subsection{Point-to-far-field lens}
\label{s7.2:Po2FFL}

\begin{figure}[!ht]
  \centering
\begin{subfigure}[b]{0.45\textwidth}
    \includegraphics[width = 1.0\textwidth,trim={0cm 0cm 0cm 0cm},clip ]{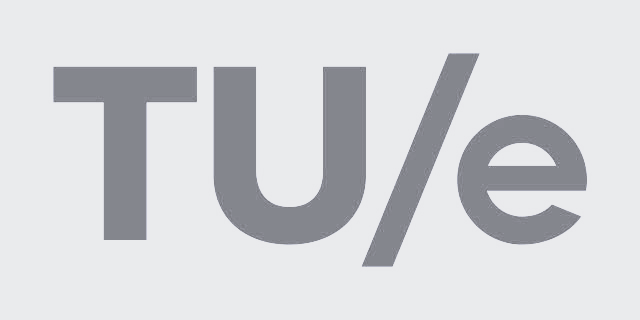}
    \vspace*{0.8cm}
    \caption{ TU/e-logo as target intensity pattern. }
    \label{Fig:TUElogo}
\end{subfigure}
\hfill
\begin{subfigure}[b]{0.45\textwidth}
    \includegraphics[width = 1.0\textwidth,trim={0cm 0cm 0cm 0cm},clip ]{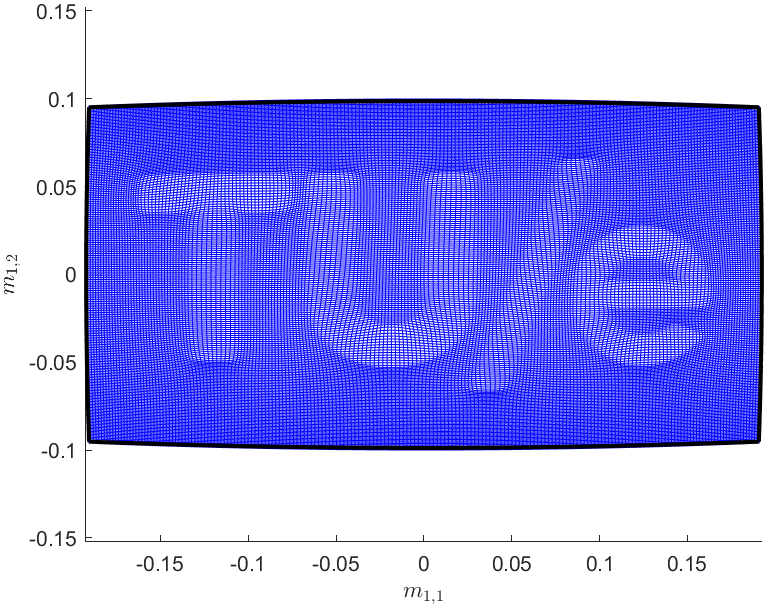}
    \caption{ Mapping. }
    \label{Fig:TUEmapping}
\end{subfigure}
\begin{subfigure}[b]{0.5\textwidth}
    \includegraphics[width = 1.0\textwidth,trim={0cm 0cm 0cm 0cm},clip ]{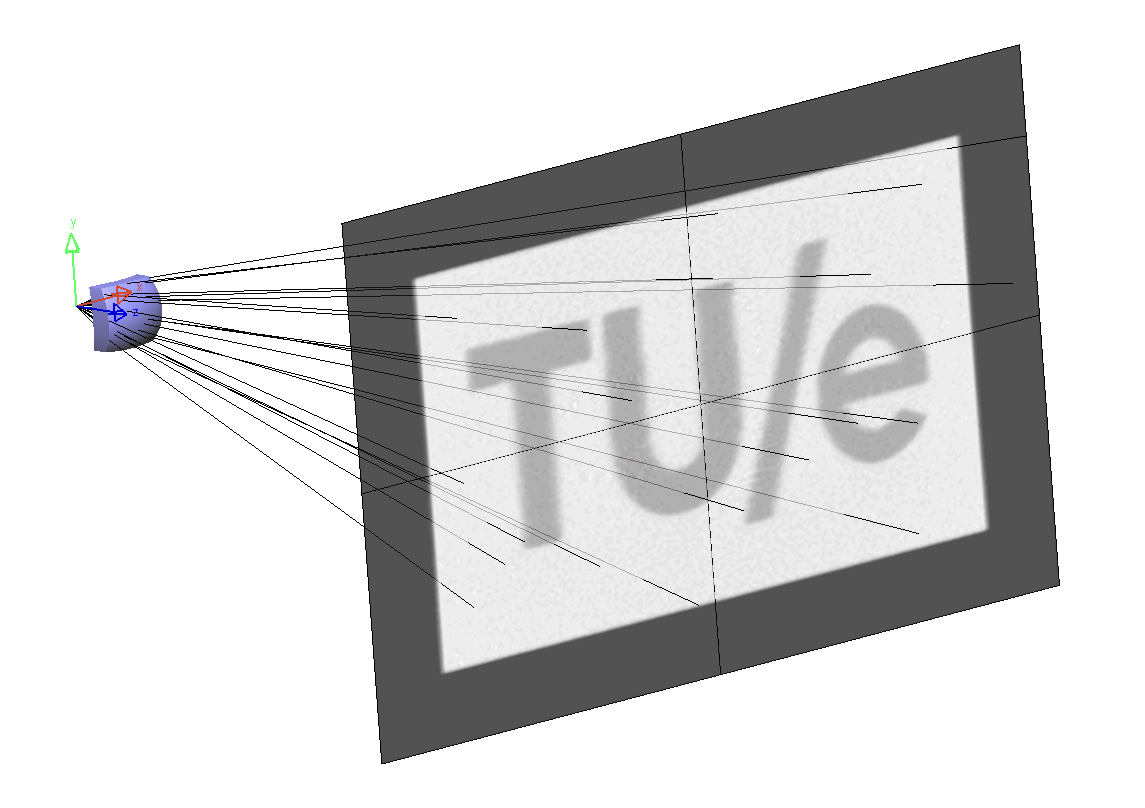}
    \caption{ Freeform lens and ray-traced target intensity pattern. }
    \label{Fig:TUElens}
\end{subfigure}
\hfill
\begin{subfigure}[b]{0.4\textwidth}
    \includegraphics[width = 1.0\textwidth,trim={0cm 0cm 0cm 0cm},clip ]{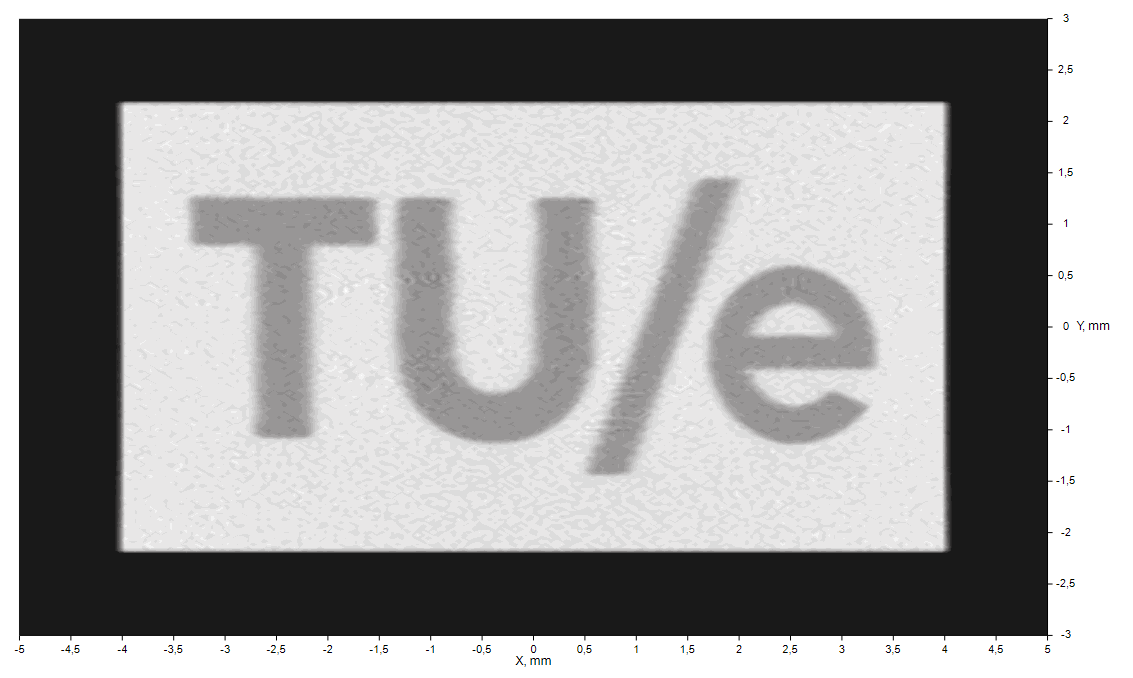}
    \vspace*{1cm}
    \caption{ Ray-traced target intensity pattern. }
    \label{Fig:TUEtarget}
\end{subfigure}
\caption{ Point-to-far-field lens: target intensity pattern, mapping, freeform surface and ray-traced target of TU/e-logo. }
\label{Fig:Po2FFL}
\end{figure}

\noindent
We consider a point source, located in the origin $\rmO_{\rms}$ of the  source plane $z = 0$, emitting upwards a
beam of light with uniform intensity $I_{\rms}(\hat{\bss}(\bsx))$ on the circular stereographic source domain $\mcX = \{\bsx \in \dR^{2} \big| |\bsx| < 0.2 \}$. In terms of spherical coordinates $(\phi,\theta)$, this corresponds to the domain $0 \le \phi \le \arccos(12/13)$ and $0 \le \theta < 2 \pi$  with $\phi$ and $\theta$ the polar and azimuthal angle, respectively, i.e., the emitted conical light beam has an opening angle of $\arccos(12/13) = 0.39\, \mathrm{rad}$. For the target we choose the stereographic domain $\mcY = [-0.2,0.2] \times [-0.1,0.1]$ and intensity $I_{\rmt}(\hat{\bst}(\bsy))$ corresponding to the gray-scale image of the TU/e logo given in Figure \ref{Fig:TUElogo}. Both intensities are scaled such that the global luminous flux balance holds, i.e., equation (\ref{s5:flux_balance31}) for $\mcA = \mcX$ and $\bsm(\mcA) = \mcY$. Both stereographic coordinates are projections from the south pole; cf. (\ref{s4:projection_south}). The refractive index of the lens $n = 1.5$ and we enforce a unique solution for the lens surface by setting $u(\bsx_{\rmc}) = 1$, and therefore $u_{1}(\bsx_{\rmc}) = 0$, corresponding to the central light ray with stereographic source coordinate $\bsx_{\rmc} = \bzero$ and direction vector $\hat{\bss}(\bsx_{\rmc}) = \hat{\bse}_{z}$; cf. Eq. (\ref{s3:opt_transport_a}).

For space discretization of (\ref{s6:BVP_m}) and (\ref{s6:BVP_u1}) we cover the source domain with a uniform $201 \times 201$ grid and evaluate $500$ iterations of the scheme (\ref{s6:iter}) - (\ref{s6:spaces}), where the associated functionals are defined in (\ref{s6:functionals}). We choose $\alpha = 10^{-3}$. The optical mapping, computed as image of the source grid is shown in Figure \ref{Fig:TUEmapping}, and clearly shows the TU/e-logo. The computed freeform lens is shown in Figure \ref{Fig:TUElens}. To validate our result, we have computed a ray-traced target intensity for this lens with $5 \times 10^{6}$ rays using the commercial software code LightTools. The results are shown in Figure \ref{Fig:TUEtarget} and Figure \ref{Fig:TUElens} and show good agreement with the desired intensity distribution in Figure \ref{Fig:TUElogo}.

\noindent
\subsection{Parallel-to-near-field reflector}
\label{s7.3:Pa2NFR}

\begin{figure}[!ht]
  \centering
\begin{subfigure}[b]{0.4\textwidth}
    \includegraphics[width = 1.0\textwidth,trim={0cm 0cm 0cm 0cm},clip ]{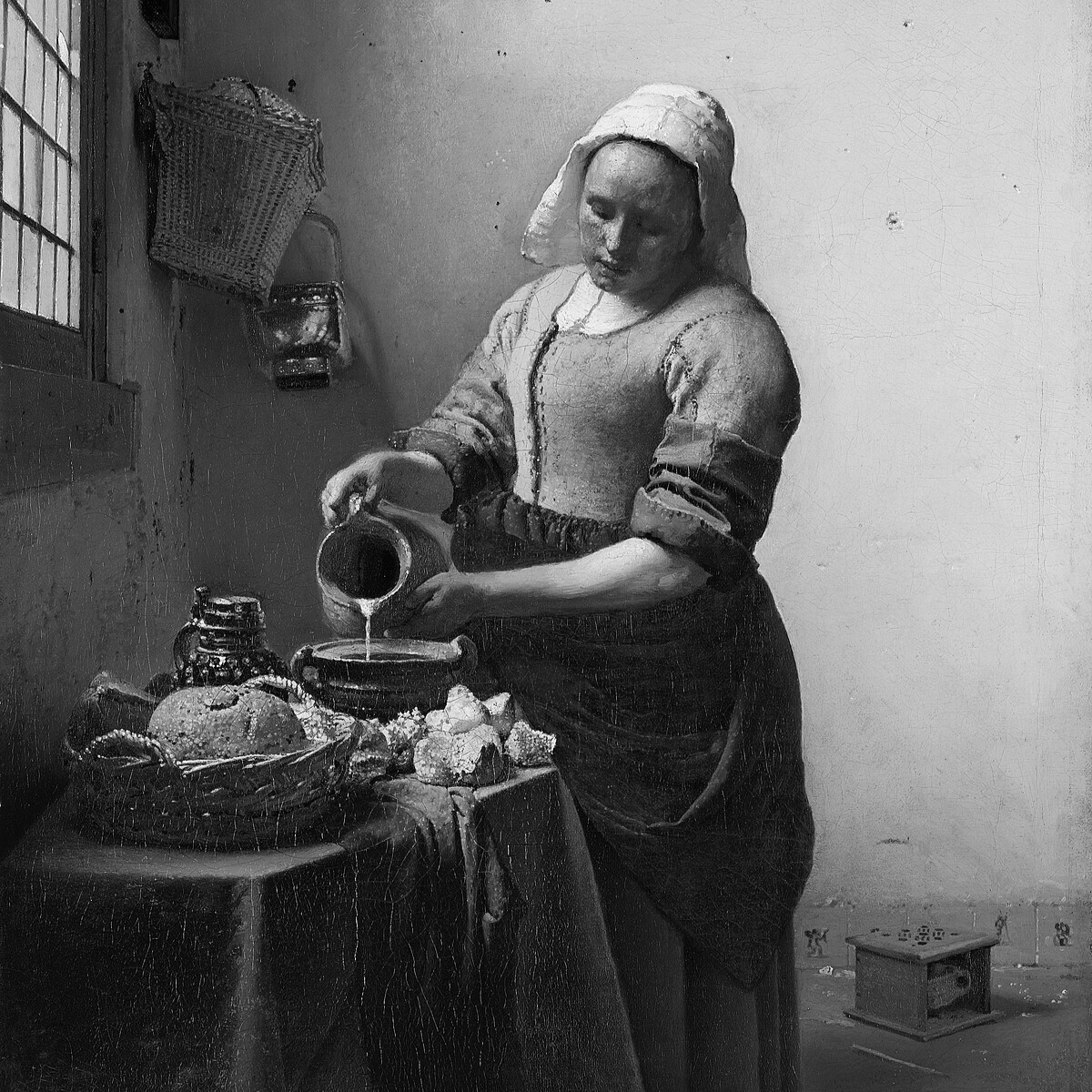}
    \caption{ `Milkmaid' as target illuminance pattern.}
    \label{fig:Milkmaid1}
\end{subfigure}
\hfill
\begin{subfigure}[b]{0.5\textwidth}
    \includegraphics[width = 1.0\textwidth,trim={0cm 0cm 0cm 0cm},clip ]{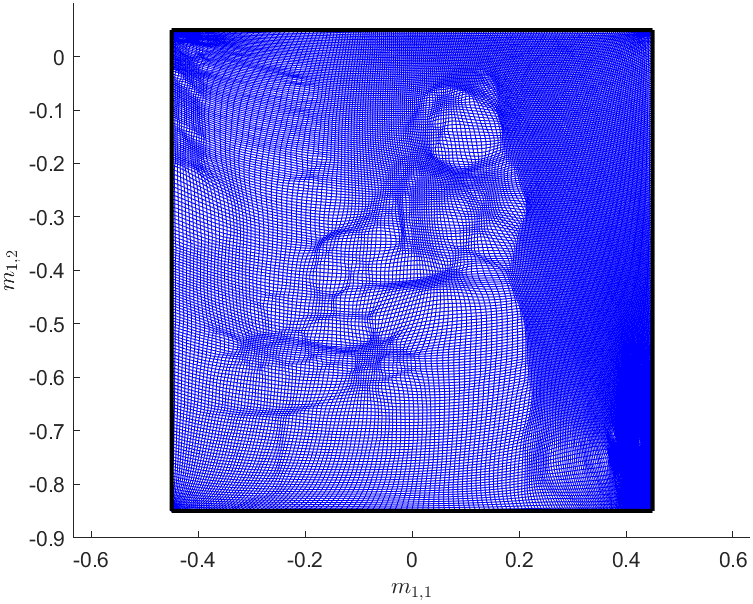}
    \caption{ Mapping. }
    \label{fig:Milkmaid2}
\end{subfigure}
\begin{subfigure}[b]{0.55\textwidth}
    \includegraphics[width = 1.0\textwidth,trim={0cm 0cm 0cm 0cm},clip ]{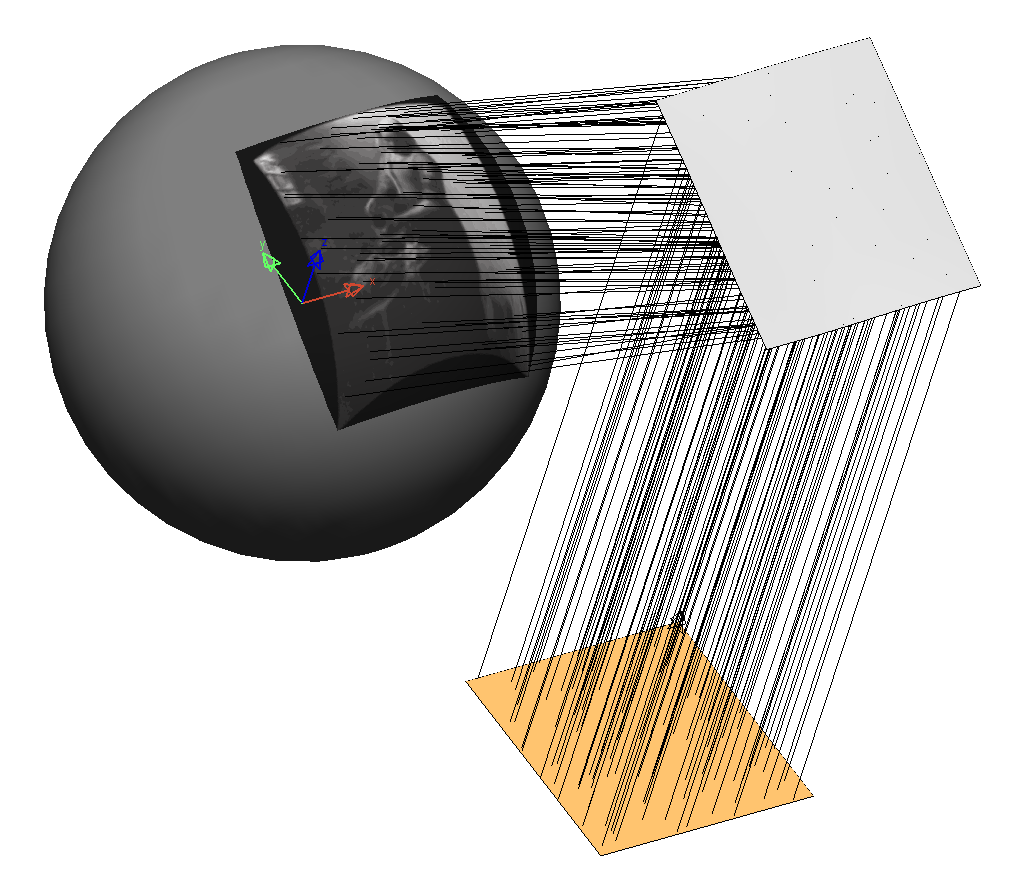}
    \caption{ Freeform reflector and ray-traced target illuminance pattern on a sphere. }
    \label{fig:Milkmaid4}
\end{subfigure}
\hfill
\begin{subfigure}[b]{0.4\textwidth}
    \includegraphics[width = 1.0\textwidth,trim={0cm 0cm 0cm 0cm},clip ]{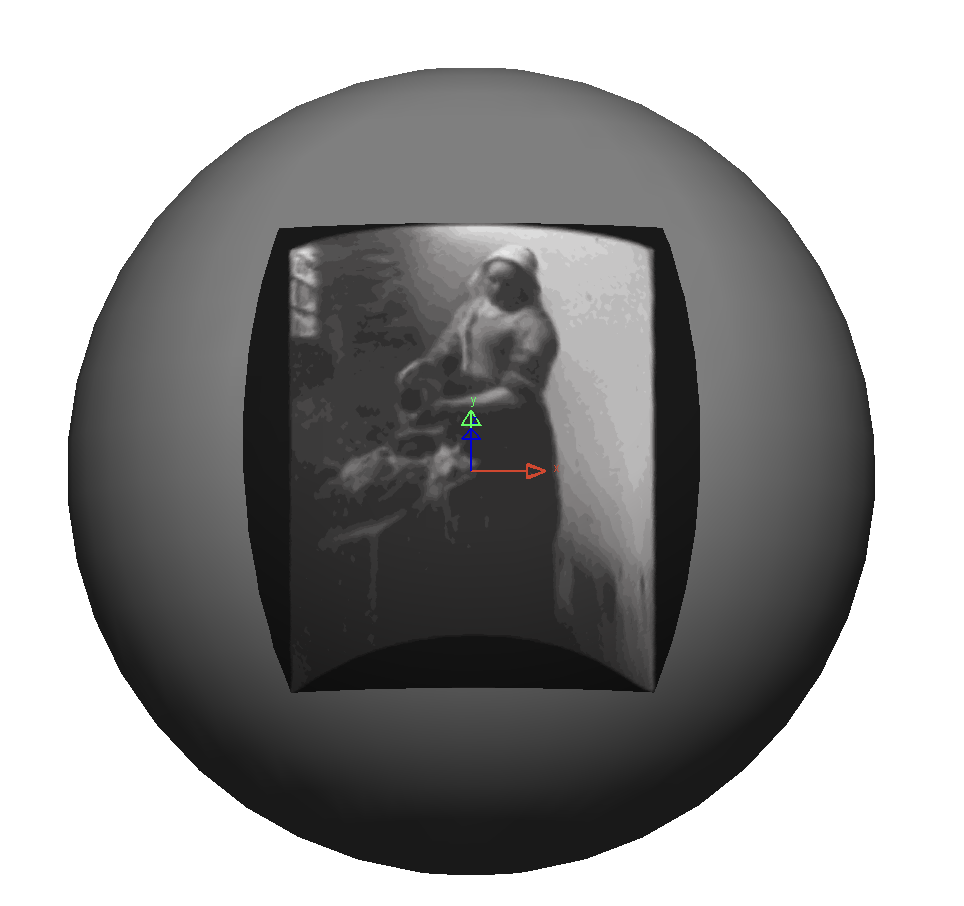}
    \vspace*{1cm}
    \caption{ Zoomed-in picture of ray-traced target illuminance pattern. }
    \label{fig:Milkmaid5}
\end{subfigure}
\caption{ Parallel-to-near-field reflector: target illuminance pattern, mapping, freeform surface and ray-traced target of the `Milkmaid'. }
\label{fig:Milkmaid}
\end{figure}

\noindent
We consider a reflector that converts a uniform source emittance $M(\bsx,0)$ into the illuminance $E(\bsr(\bsy))$ corresponding to the gray-scale image of the painting 'the Milkmaid' by Johannes Vermeer as shown in Figure \ref{fig:Milkmaid1}, projected on a spherical target surface. The source domain $\mcX = [ -0.5,0.5 ] \times [ -3,-2 ]$ and the target domain is located on the northern hemisphere of the unit sphere given by $z = v(\bsy) = \sqrt{ 1 - |\bsy|^{2} }$ with $\mcY = [ -0.42,0.42 ] \times [-0.82,0.02 ]$. The parameters $\bsx$ and $\bsy$ are both Cartesian coordinates. The emittance and illuminance are scaled such that the global flux balance holds, i.e., Eq. (\ref{s5:balance4}) for $\mcA = \mcX$ and $\bsm(\mcA) = \mcY$. We enforce a unique solution for the location of the reflector by setting $u(\bsx_{\rmc}) = u_{1}(\bsx_{\rmc}) = 3$ where $\bsx_{\rmc} = (0,-2.5)^{\rmT}$ is the center point of the source domain $\mcX$.

To discretize the BVPs (\ref{s6:BVP_m}) and (\ref{s6:BVP_u11}) we cover the source domain $\mcX$ with a uniform $201 \times 201$ grid and apply the iteration scheme (\ref{s6:iter1}), (\ref{s6:spaces}) $500$ times.
The associated functionals are defined in (\ref{s6:functionals}), where we choose $\alpha = 0.05$, and  (\ref{s6:funcI}). The optical map as image of the uniform rectangular source grid is shown in Figure \ref{fig:Milkmaid2}. Clearly, the girl shown in Figure \ref{fig:Milkmaid1} is recognizable in the optical mapping, as grid points come closer together in regions of high contrast showing her contours.

To verify the reflector computed by the least-squares algorithm, we used LightTools. Implementing the reflector coordinates into this code and specifying the uniformly distributed rectangular source along with the curved target surface, we obtained the reflector system illustrated in Figure \ref{fig:Milkmaid4}, where only $50$ rays are shown. The resulting target intensity, which can also be seen in the zoomed-in picture of Figure \ref{fig:Milkmaid5}, was obtained from tracing $5 \times 10^{7}$ rays. The image of the `Milkmaid' is clearly visible, confirming the validity of the reflector computed by the least-squares algorithm.



\section{Concluding remarks}
\label{s8:Conclusions}

We have presented a systematic and generic derivation of the governing equations for inverse optical design. Using Hamilton's characteristic functions, we could derive a geometrical equation defining the shape/location of the optical surface(s). From this equation, applying concepts from optimal transport theory, we derived equations for the optical map. To close the model, we presented a generic conservation law of luminous flux. Combining all equations, we derived a Jacobian equation for the optical map. Subsequently, we elaborated the generic model in three different models of increasing complexity, on the basis of three example optical systems. We briefly outlined numerical least-squares methods for all models, and demonstrated their performance for a few examples.

We intend to expand our research on inverse methods along the following lines. First, we will adjust our numerical methods to incorporate two target distributions, second, we will develop mathematical models and numerical methods for catadioptic systems where light rays propagate via different paths from source to target, and finally, and probably the most challenging, we intent to generalize our models and numerics to optical systems with finite sources as opposed to zero-\'{e}tendue sources.

\section*{Nomenclature}

\noindent
\textbf{Scalar variables}\\[0.2cm]
\begin{tabular}{ll}
  $c(\bsx,\bsy)$                          & cost function\\
  $d(\rmP,\rmQ)$                          & distance between the points $\rmP$ and $\rmQ$\\
  $d_{\rm opt}(\bsr_{\rms},\bsr_{\rmt})$  & optical distance, i.e., optical path length $[\rmP_{\rms},\rmP_{\rmt}]$ as function of the position vectors $\bsr_{\rms}$ and $\bsr_{\rmt}$\\
  $E(\bsr(\bsy))$                         & illuminance at a near-field target\\
  $f(\bsx)$                               & generic flux density of source\\
  $g(\bsy)$                               & generic flux density of target\\
  $G(\bsx,\bsy,w)$                        & generating function\\
  $H(\bsx,\bsy,w)$                        & inverse of generating function\\
  $I_{\rms}(\hat{\bss}(\bsx))$            & intensity of point source\\
  $I_{\rmt}(\hat{\bst}(\bsy))$            & intensity of far-field target\\
  $J(\hat{\bsv}(\bsz))$                   & factor in the area element $\rmd S(\bsz)$ on $\rmS^{2}$ generated by $\hat{\bsv}$ and parameterized by $\bsz$\\
  $M(\bsr(\bsx))$                         & emittance of planar source\\
  $n(\bsr)$                               & refractive index field\\
  $s$                                     & arc length along a curve $\mcC$ representing a ray\\
  $T(\bsp_{\rms},\bsp_{\rmt})$            & angular characteristic function\\
  $u_{1}(\bsx)$                           & Kantorovich potential for source domain\\
  $u_{2}(\bsy)$                           & Kantorovich potential for target domain\\
  $V(\bsq_{\rms},\bsq_{\rmt})$            & point characteristic function\\
  $W(\bsq_{\rms},\bsp_{\rmt})$            & mixed characteristic function of the first kind\\
  $W^{\ast}(\bsp_{\rms},\bsq_{\rmt})$     & mixed characteristic function of the second kind\\
  $z$                                     & coordinate along the optical axis\\
  $\varphi(\bsr)$                         & phase of an electromagnetic wave\\
  $[\rmP_{\rms},\rmP_{\rmt}]$             & optical path length between the points $\rmP_{\!\rms}$ and $\rmP_{\!\rmt}$
\end{tabular}
\vspace{0.1cm}\\

\clearpage
\noindent\textbf{Vectors}\\[0.2cm]
\begin{tabular}{ll}
  $\bsp(z)$               & 2D momentum vector of a ray\\
  $\bsq(z)$               & 2D position vector of a ray\\
  $\bsr$                  & 3D position vector of an arbitrary point\\
  $\hat{\bss}$            & 3D unit direction/tangent vector of a ray at the source plane\\
  $\hat{\bst}$            & 3D unit direction/tangent vector of a ray at the target plane\\
  $\hat{\bsv}$            & 3D unit direction/tangent vector at an arbitrary point of a ray\\
  $\bsx$                  & 2D vector parameterizing the source domain\\
  $\bsy$                  & 2D vector parameterizing the target domain
\end{tabular}


\section*{Conflict of Interest Statement}

The authors declare that the research was conducted in the absence of any commercial or financial relationships that could be construed as a potential conflict of interest.


\section*{Funding}

This work was partially supported by the Dutch Research Council (\textit{Dutch}: Nederlandse Organisatie voor Wetenschappelijk Onderzoek -- NWO) through the grants P15-36 and P21-20.

\end{document}